\documentclass[aps,prd,twocolumn,eqsecnum,showpacs,preprintnumbers]{revtex4}
\usepackage{graphicx}

\newcommand{\be}{\begin{equation}}
\newcommand{\ee}{\end{equation}}
\newcommand{\ba}{\begin{eqnarray}}
\newcommand{\ea}{\end{eqnarray}}
\newcommand{\ep}{\varepsilon}
\newcommand{\nn}{\nonumber}
\newcommand{\ra}{\rightarrow}
\newcommand{\lra}{\leftrightarrow}

\newcommand{\ita}{\textit}

\newcommand{\nc}{N_C}

\newcommand{\hspn}{{\hspace{-0.6mm}}}
\newcommand{\cf}{{C^{}_F}}

\begin{document}

\preprint{DESY 08--131}

\title{Heavy quark pair production in gluon fusion at next-to-next-to-leading 
${\cal O}(\alpha_s^4)$ order: One-loop squared contributions}



\author{B.\ A.\ Kniehl}
\email{kniehl@desy.de}
\affiliation{ II. Institut f\"{u}r Theoretische Physik,
Universit\"{a}t Hamburg, Luruper Chaussee 149, 22761 Hamburg, Germany}

\author{J.\ G.\ K\"{o}rner}
\email{koerner@thep.physik.uni-mainz.de}
\affiliation{Institut f\"{u}r Physik, Johannes
Gutenberg-Universit\"{a}t, 55099 Mainz, Germany}

\author{Z.\ Merebashvili}
\email{zakaria.merebashvili@desy.de}
\affiliation{ II. Institut f\"{u}r Theoretische Physik,
Universit\"{a}t Hamburg, Luruper Chaussee 149, 22761 Hamburg, Germany}

\author{M.\ Rogal}
\email{Mikhail.Rogal@desy.de}
\affiliation{Deutsches Elektronen-Synchrotron DESY, Platanenallee 6, 15738
Zeuthen, Germany}

\date{\today}

\begin{abstract}
We calculate the next-to-next-to-leading-order ${\cal O}(\alpha_s^4)$
one-loop squared corrections to the production
of heavy-quark pairs in the gluon-gluon fusion process.
Together with the previously derived results on the $q \bar{q}$ production
channel, the results of this paper complete the calculation of the one-loop 
squared contributions of the next-to-next-to-leading-order
${\cal O}(\alpha_s^4)$ radiative QCD
corrections to the hadroproduction of heavy flavors.
Our results, with the full mass dependence retained,
are presented in a closed and very compact form, in dimensional
regularization.
\end{abstract}

\pacs{12.38.Bx, 13.85.-t, 13.85.Fb, 13.88.+e}

\maketitle

\section{\label{intro}Introduction}

It has been already 20 years since the next-to-leading-order (NLO)
corrections to the hadroproduction of heavy flavors were first presented in
the seminal work \cite{Ellis}. These 
results were confirmed yet in another
seminal work \cite{Been}.

In the past few years there was much progress in describing the 
experimental results on heavy-flavor production.
For instance, in a recent work 
\cite{Bernd} it was shown that a NLO analysis of the transverse-momentum 
distributions does
in fact properly describe the latest bottom quark
production data \cite{CDF} in a surprisingly large kinematical range.
The improvement in the theoretical prediction is mainly due to 
advances in the analysis
of parton distribution functions and the QCD coupling constant.
We also point out the progress in dealing with numerically large mass logarithms
that spoil the convergence of the perturbative expansion in the high energy
(or small mass) asymptotic domain.
In this respect we mention the work
\cite{Bernd2} where also charm pair production is reconciled with
experimental data.
Data on top-quark
pair production also agrees with the NLO prediction within theoretical and
experimental errors
(see e.g. Ref.~\cite{Chakraborty:2003iw}).
However, in all of these NLO calculations there
remains, among others, the problem
that the renormalization and factorization scale dependences 
render the theoretical predictions to have much larger uncertanties
than today's standards require. This calls
for a next-to-next-to-leading-order (NNLO)
calculation of heavy-quark production in hadronic collisions.
In fact, the scale dependence
of the theoretical prediction is expected to be considerably reduced
when NNLO partonic amplitudes are folded with the available NNLO parton
distributions. 
For example, by approximating the NNLO 
corrections with the fixed-order expansion of the next-to-leading-log
prediction, one finds
a projected NNLO scale uncertainty of about 
3\% \cite{MochUwer}, 
which is below the parton distribution uncertainty, and in line with the 
anticipated experimental error.

Recently there was much activity in the phenomenology of hadronic 
heavy-quark pair production in connection with the Tevatron and the CERN
Large Hadron
Collider (LHC), which had its start-up this year. There will be
much experimental effort dedicated to the discovery of the Higgs boson.
There will also be studies of the copious production of top quarks and other
heavy particles, which serve as a background to Higgs
boson searches as well as to possible new physics beyond the standard
model. Therefore, it is mandatory to reduce the
theoretical uncertainty in phenomenological
calculations of heavy-quark production processes as much as possible.

\begin{figure*}
\includegraphics[height=4.0cm]{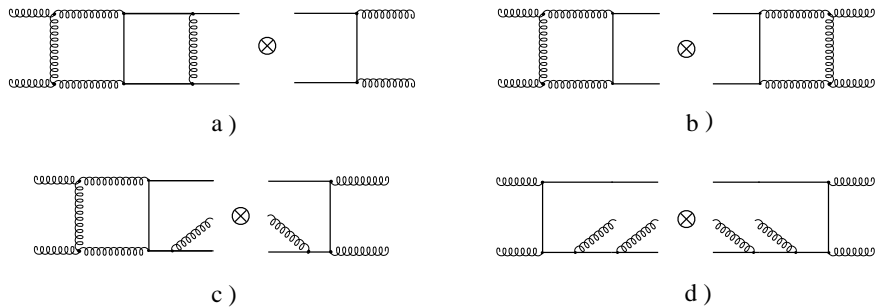}
\caption{\label{fig:exmpl} Exemplary gluon fusion diagrams for the
NNLO calculation of heavy-hadron production.}
\end{figure*}

Several years ago the NNLO contributions to
hadron production were
calculated  by several groups in massless QCD (see e.g. Ref.~\cite{Glover} and
references therein).
The
completion of a similar program for processes that involve massive
quarks requires much more dedication, since the inclusion of
an additional mass scale dramatically complicates the whole
calculation.

At the lower energies of Tevatron II, top-quark pair production is dominated
by $q {\bar q}$ annihilation (85\%). The remaining 15\% comes from
gluon fusion. At the higher energies of the LHC, gluon fusion dominates
the production process (90\%) leaving 10\% for $q {\bar q}$ annihilation
(percentage figures from Ref.~\cite{Chakraborty:2003iw}). This shows that both
$q {\bar q}$ annihilation and gluon fusion have to be accounted for in the
calculation of top-quark pair production. Since gluon fusion makes up the
largest part of the heavy-quark pair production cross section at the LHC it
is important to reduce renormalization and factorization scale uncertainties
in the gluon fusion process as much as possible in view of the fact that the
large uncertainties in the gluonic parton distribution functions translate to
large cross section uncertainties at the LHC.

There are four classes of contributions that need to be calculated for the
NNLO corrections to the hadronic production of heavy-quark pairs.
In Fig.~\ref{fig:exmpl} we show one generic diagram each for the
four classes of contributions that need to be calculated for the
NNLO corrections to the gluon-initiated hadroproduction of heavy flavors.
The first class involves
the pure two-loop contribution [\ref{fig:exmpl}(a)], which has to be
folded with the leading-order (LO) Born term.
The second class of diagrams [\ref{fig:exmpl}(b)]
consists of the so-called one-loop squared contributions (also called
loop-by-loop contributions) arising from the product of one-loop
virtual matrix elements. This is the topic of the present paper.
Further, there are the
one-loop gluon emission contributions [\ref{fig:exmpl}(c)]
that are folded with the one-gluon
emission graphs. Finally, there are the squared two-gluon emission
contributions [\ref{fig:exmpl}(d)] that are purely of tree type.
The corresponding graphs for the quark-initiated
processes are not displayed.

Bits and pieces of the NNLO calculation for hadroproduction of heavy
flavors are now being assembled.
In this context we would like to
mention the recent
two-loop calculation of the heavy-quark vertex form factor
\cite{bernreuther05a} that can be used as one of the many building blocks
in the first class of processes.
There is also a very promising numerical approach applied to the calculation
of the pure two-loop diagrams \cite{Czakon}. Recently, an analytic calculation
of a subclass of the two-loop contributions to $q \bar{q} \to Q \bar{Q}$ was 
published \cite{Bonciani:2008az}.
The authors of Ref.~\cite{Stefan} 
have calculated the NLO corrections to $t \bar{t}+$jet production with
contributions from the third class of diagrams. However, 
this result needs further subtraction terms in order to allow for an 
integration over the full phase space.
We would also like to mention the recent work on the two-loop
virtual amplitudes that are valid in the domain of high energy
asymptotics, where the heavy-quark mass is small compared to the other large
scales.
In this calculation \cite{Moch}, mass power
corrections are left out, and only large mass logarithms and finite terms
associated with them are retained.
Much work was also done in relation to the resummation of soft 
contributions.
In this respect we refer the reader to recent publications 
where some different
approaches to the resummation are advocated \cite{MochUwer,soft}.

\begin{figure*}
\includegraphics{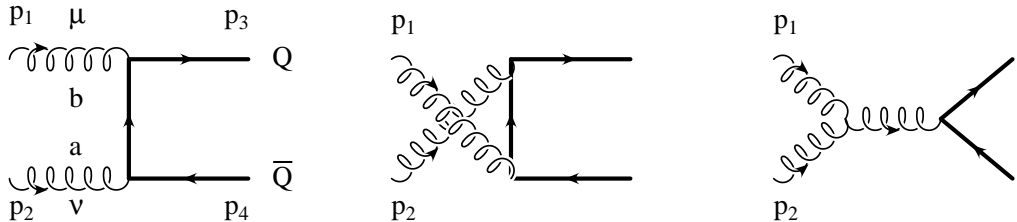}
\caption{\label{fig:born}
The $t$-, $u$-, and $s$-channel LO graphs contributing
to the gluon (curly lines) fusion amplitude.
The thick solid lines correspond to the heavy quarks.}
\end{figure*}

The authors of the
present paper have been involved in a systematic effort
to calculate all the contributions from the second class of processes, i.e. the
one-loop squared contributions. The NNLO one-loop squared amplitudes
for the quark-initiated process were recently presented in Ref.~\cite{KMRqq}.
In this paper, we report on a calculation of the NNLO one-loop
squared matrix elements for the process
$gg \to Q \overline{Q}$. The
calculation is carried out in dimensional
regularization \cite{DREG} with space-time dimension $n=4-2\ep$. We 
mention that we have presented closed-form, one-loop squared results for 
heavy-quark production in the fusion of real photons in Ref.~\cite{gamgam}. 
With the present paper the program of calculating the one-loop squared 
contributions to heavy-quark pair hadroproduction has now been completed. 

Let us briefly describe some of the main features of the calculation of the
one-loop squared contributions.
The highest singularity in the one-loop amplitudes arises from infrared (IR) and
mass singularities (M) and is thus, in general, proportional to $(1/\ep^2)$. This in turn
implies that the Laurent series expansion of the one-loop amplitudes
has to be taken up to ${\cal O}(\ep^2)$ when calculating the one-loop squared
contributions. In fact, it is the ${\cal O}(\ep^2)$ terms in the Laurent
series expansion that really complicate things \cite{KMR}, since the
${\cal O}(\ep^2)$ contributions in the one-loop amplitudes involve a
multitude of multiple polylogarithms of maximal weight and depth 4
\cite{KMR1}. All scalar master integrals needed in this calculation have
been assembled in Refs.~\cite{KMR,KMR1}. Reference \cite{KMR} gives 
the results in terms 
of so-called $L$ functions, which can be written as one-dimensional integral
representations involving products of log
and dilog functions, while Ref.~\cite{KMR1} gives the results in terms of multiple
polylogarithms. The divergent and finite terms of the one-loop {\it amplitude}
for $gg \to Q{\overline Q}$ were given in Ref.~\cite{KM}. The remaining
${\mathcal O}(\ep)$ and ${\mathcal O}(\ep^{2})$ amplitudes have
been written down in Ref.~\cite{KMR2}. We shall rewrite these matrix elements in a
representation more suitable for the purposes of the present application.

In our presentation, we shall
make use of our notation for the coefficient functions of the relevant
scalar one-loop master integrals calculated up to ${\cal O}(\ep^2)$ in
Refs.~\cite{KMR,KMR1}.
For the case of gluon-gluon and quark-antiquark collisions, one needs
all the scalar integrals derived in Refs.~\cite{KMR,KMR1}, e.g.
the one scalar one-point
function $A$, the five scalar two-point functions $B_1$, $B_2$, $B_3$,
$B_4$, and $B_5$, the six scalar
three-point functions $C_1, C_2, C_3, C_4, C_5$, and $C_6$, and three
scalar four-point
functions $D_1, D_2$, and $D_3$.
Taking the {\it complex} scalar four-point function $D_2$ as an example,
we define successive coefficient functions $D_2^{(j)}$ for the Laurent
series expansion of $D_2$.
One has
\begin{eqnarray}
\label{Dexp}
D_2&\!\!=\!\!&i C_\ep(m^2)\Big\{\frac{1}{\ep^{2}}D_2^{(-2)}+\frac{1}{\ep}D_2^{(-1)} +
D_2^{(0)} + \ep D_2^{(1)} \nonumber \\
&& + \ep^2 D_2^{(2)} + {\mathcal O}(\ep^3) \Big\},
\end{eqnarray}
where $C_{\ep}(m^2)$ is defined by
\be
\label{ceps}
C_{\ep}(m^2)\equiv\frac{\Gamma(1+\ep)}{(4\pi)^2}
\left(\frac{4\pi\mu^2}{m^2}\right)^\ep .
\ee
We use this notation for both the real and imaginary parts of $D_2$,
i.e. for ${\rm Re}D_2$ and ${\rm Im}D_2$.
Similar expansions hold for the scalar one-point
function $A$, the scalar
two-point functions $B_i$, the scalar three-point functions $C_i$,
and the remaining four-point functions $D_{i}$.
The coefficient functions of the various Laurent
series expansions were given in Ref.~\cite{KMR} in the form of so-called
$L$ functions, and in Ref.~\cite{KMR1} in terms of multiple
polylogarithms of maximal weight and depth 4. It is then a matter of
choice which of the two representations are used for the numerical
evaluation. The numerical evaluation of the $L$ functions in terms of their
one-dimensional integral representations is quite straightforward using
conventional integration routines, while there exists a very efficient
algorithm to numerically evaluate multiple polylogarithms
\cite{Vollinga:2004sn}.

Let us briefly summarize the main features of the scalar master integrals. The master
integrals $A,B_{1},B_{3},B_{4},C_{2},C_{3}$, and $D_{3}$ are real, whereas
$B_{2},B_{5},C_{1},C_{4},C_{5},C_{6},D_{1}$, and $D_{2}$ are complex. From
the form $(AB^{\ast}+BA^{\ast})=2({\rm Re}A\,{\rm Re}B+{\rm Im}A\,{\rm Im}B)$
it is clear that the imaginary parts of the master integrals
must be taken into account in the one-loop squared contribution. The master
integrals
$B_{2},B_{5},C_{1},C_{4},C_{5}$, and $C_{6}$ are $(t\lra u)$ symmetric,
where the kinematic variables $t$ and $u$ are defined in Sec.~\ref{notation}.

This paper is organized as follows. Section.~\ref{notation} contains an
outline of our general approach and discusses renormalization procedures.
Section.~\ref{nlo} presents LO and NLO results
for the gluon fusion subprocess.
In Sec.~\ref{singular} one finds a discussion of the singularity
structure of the NNLO squared matrix element for the gluon fusion
subprocess.
In Sec.~\ref{finite} we discuss the structure of the finite part of our
result.
Our results are summarized in Sec.~\ref{summary}.
In the Appendices, we present expressions for various coefficients 
that are used in Sec.~\ref{nlo} to write down the NLO result.


\section{\label{notation}
NOTATION and renormalization
}


Heavy-flavor hadroproduction proceeds through two partonic
subprocesses: gluon fusion and light-quark-antiquark annihilation. The
first subprocess is the most challenging one in QCD from a technical point
of view. It has three production topologies already at the Born level
(see Fig.~\ref{fig:born}).
The second subprocess, where there is only one topology
at the Born level, was considered in Ref.~\cite{KMRqq}. 
Irrespective of the partons
involved, the general kinematics is, of course, the same in both processes. In
particular, for gluon fusion, Fig.~\ref{fig:born}, we have
\be
\label{gg}
g(p_1) + g(p_2) \ra Q(p_3) + \overline Q(p_4),
\ee

The momentum flow directions correspond to the physical configuration, e.g.
$p_1$ and
$p_2$ are ingoing whereas $p_3$ and $p_4$ are outgoing.
With $m$ being the heavy-quark mass, we define
\ba
\nn
&s\equiv (p_1+p_2)^2, \qquad  t\equiv T-m^2 \equiv
(p_1-p_3)^2-m^2,&
\\
&u\equiv U-m^2\equiv (p_2-p_3)^2-m^2, &
\ea
so that one has the energy-momentum conservation relation $s+t+u=0$.

We also introduce the overall factor
\be
\label{common}
\mathcal C = \left( g_s^4 C_{\ep}(m^2) \right)^2,
\ee
where $g_s$ is the renormalized strong-coupling constant
and $C_\ep(m^2)$ is defined in Eq.~(\ref{ceps}).

As was shown e.g. in Refs.~\cite{KM,KMR2} the self-energy
and vertex diagrams contain ultraviolet (UV), infrared and collinear
(IR/M) poles after
heavy-mass renormalization. The UV poles need to be regularized.

Our renormalization procedure is carried out in a mixed
renormalization scheme. When dealing with massless
quarks, we work in the modified minimal-subtraction 
($\overline{\rm MS}$) scheme, while heavy quarks are
renormalized in the on-shell scheme  defined by the following
conditions for the renormalized external heavy-quark self-energy graphs:
\be
\label{OS}
\Sigma_{r}(\not p)|_{\not p=m} = 0,
\qquad
\frac{\partial}{\partial{\rm \hspace{-.07in}}\not p} \Sigma_{r}(\not
p)|_{\not p=m} = 0.
\ee
In the on-shell scheme, the first condition in Eq.~(\ref{OS}) ensures 
that the heavy-quark mass is the pole mass.

For completeness, we list the set of one-loop renormalization constants 
used in this paper. One has
\ba
\label{renormconstants}
\nn   &&
Z_1 = 1 + \frac{g^2_s}{\ep} \frac{2}{3} \left\{ (N_C - n_{l}) C_{\ep}(\mu^2)
                   - C_{\ep}(m^2) \right\},     \\
\nn   &&
Z_m = 1 - g^2_s C_F C_{\ep}(m^2) \frac{3-2\ep}{\ep (1-2\ep)},    \\
&&
Z_2 = Z_m,   \\
\nn   &&
Z_{1F} = Z_2 - \frac{g^2_s}{\ep} N_C C_{\ep}(\mu^2),    \\
\nn   &&
Z_{1f} = 1 - \frac{g^2_s}{\ep} N_C C_{\ep}(\mu^2),    \\
\nn   &&
Z_3 = 1 + \frac{g^2_s}{\ep} \left\{ (\frac{5}{3} N_C - \frac{2}{3}
n_{l}) C_{\ep}(\mu^2) - \frac{2}{3} C_{\ep}(m^2)\right\}
\\  \nn &&  \qquad
= 1 + \frac{g^2_s}{\ep}\left\{ (\beta_0 - 2N_C) C_{\ep}(\mu^2) -
\frac{2}{3} C_{\ep}(m^2)\right\},      \\
\nn   &&
Z_g =  1 - \frac{g^2_s}{\ep}\left\{ \frac{\beta_0}{2} C_{\ep}(\mu^2) -
\frac{1}{3} C_{\ep}(m^2)\right\},
\ea
with $\beta_0=(11 N_C - 2 n_{l})/3$ being the first coefficient of the 
QCD beta function, 
$n_{l}$ the number of light quarks,
$C_F=4/3$, and $N_C=3$ the number of colors. The arbitrary mass
scale $\mu$ is the scale at which the renormalization is carried out. The above
renormalization constants renormalize the following quantities: $Z_1$ for the
three-gluon vertex,
$Z_m$ for the heavy-quark mass, $Z_2$ for the heavy-quark wave function,
$Z_{1F}$ for the $(Q\overline Qg)$ vertex,
$Z_{1f}$ for the $(q\overline qg)$ vertex,
$Z_3$ for the gluon wave function and $Z_g$ for the strong-coupling
constant $\alpha_{s}$.
For the massless quarks, there is no mass and wave function renormalization.

\begin{figure*}
\includegraphics[height=15.0cm]{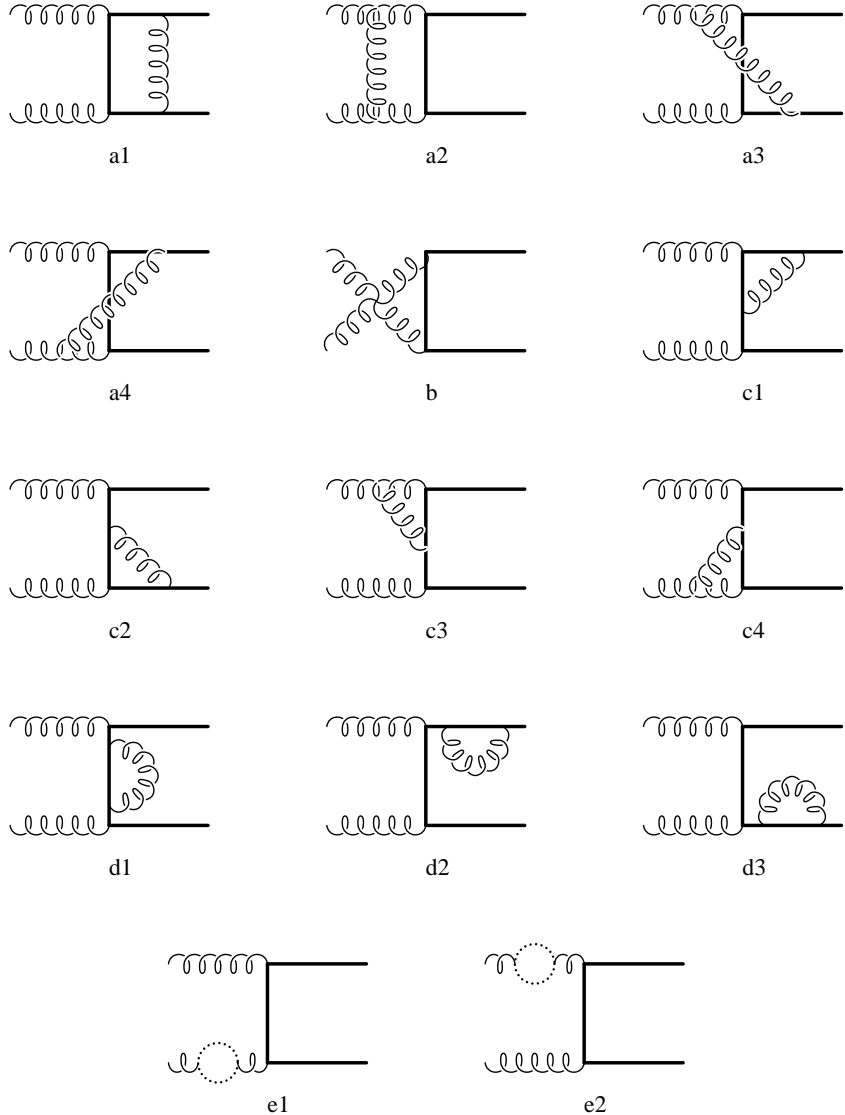}
\caption{\label{fig:ggnlot}
The $t$-channel one-loop graphs contributing to the gluon fusion amplitude.
Loops with dotted lines represent the gluon, ghost, and light and
heavy quarks.}
\end{figure*}

Let us sketch
the two alternative ways of getting the final one-loop-renormalized
amplitude from the mass-renormalized amplitude:

\noindent
i) Take the given mass-renormalized matrix element or the square of that
matrix element and
multiply all the self-energy graphs by a factor 1/2. 
Then renormalize the coupling constant in the LO Born amplitude.

\noindent
ii) Take the given mass-renormalized matrix element and apply the corresponding
counterterms obtained from the LO matrix element by inserting the relevant
$Z-1$ factors into the {\it internal} propagators and vertices.
All the renormalization constants we need are presented in 
Eq.~(\ref{renormconstants}).
We will get the
renormalized vertex function $\Gamma_R^{(N)}$, where $(N)$ denotes the set of
$N$ external particles.
The renormalized matrix element is obtained from
\be
\label{renormalized}
M_R = \Gamma_R^{(N)} \prod_{i=1}^N \left( Z_R^{(i)} \right)^{\frac{1}{2}},
\ee
where $Z_R^{(i)}$ are the residues of the renormalized propagators at the poles
for all the particles under consideration. They are related to the residues of the
unrenormalized propagators via
\be
\label{relation}
Z_R^{(i)} = Z_U^{(i)} Z_i^{-1}
\ee
where the $Z_i$
are the respective external wave function renormalization constants.

Working at the one-loop order,
we note that in the on-shell scheme $Z_R^{(i)}=1$. This is a direct consequence of the
second condition in Eq.~(\ref{OS}), which effectively cuts off the external massive 
lines.
For the case of external massless partons $Z_U^{(i)}=1$.
It is important to note that the
gluon wave function renormalization constant $Z_3$ is a mixture of two parts:
the part which multiplies $C_{\ep}(\mu^2)$ is derived in the 
$\rm \overline{MS}$ scheme,
while the last term due to the heavy-quark loop is derived in the on-shell scheme.
For this reason, this last term has to be omitted in $Z_3$ when using 
it as an external
field renormalization constant in Eq.~(\ref{relation}).
Since in our case we have two gluon and two heavy-quark fields, we therefore obtain
\be
M_R = \Gamma_R^{(N)} Z_3^{-1}.
\ee

The final result should not depend on which of the two ways has been chosen to
do the renormalization. We have
checked that, in both ways, one arrives at the same renormalized matrix element.

In order to fix our normalization, we write down the differential cross section
for $g g \to Q \overline{Q}$ in
terms of the squared amplitudes $|M|^2$.
One has
\be
d\sigma_{g g \rightarrow  Q \overline{Q} }=
\frac{1}{2s}
\frac{d({\rm PS})_2}{4 (1-\ep)^2} \frac{1}{d_A^2}
|M|^2_{g g \rightarrow  Q \overline{Q} } \, ,
\ee
where the $n$--dimensional two--body phase space is given by
\be
d ({\rm PS})_2=
\frac{m^{-2\ep}}{8 \pi s} \frac{(4\pi)^{\ep}}{\Gamma (1-\ep)}
\left( \frac{tu-sm^2}{sm^2} \right)^{-\ep} \delta (s+t+u) dtdu\,.
\ee
We explicitly exhibit the flux factor $(4p_1p_2)^{-1} = (2s)^{-1}$, and
the spin $(n-2)^{-2}=(2-2\ep)^{-2}$ and color $d_{A}^{-2}$
averaging factors for the initial gluons.
Here $d_A=N_C^2-1=8$ is the dimension of the adjoint
representation of the color group SU($N_C$).

\begin{figure*}
\includegraphics[height=9.0cm]{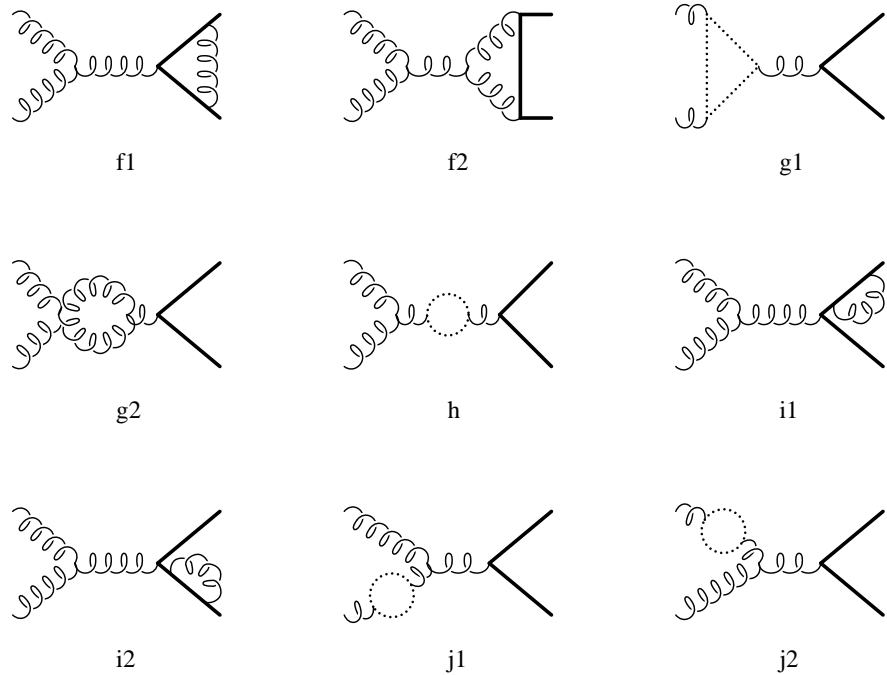}
\caption{\label{fig:ggnlos}
The $s$-channel one-loop graphs contributing to the gluon fusion amplitude.
Loops with the dotted lines as in g1, h, j1, and j2 represent the gluon,
ghost, and light and
heavy quarks. The four-gluon coupling contribution appears in g2.}
\end{figure*}


\section{\label{nlo}
Leading and Next-to-leading order results
}


At LO for 
$g g \rightarrow  Q \overline{Q}$,
we shall use a representation which differs from the one given in 
Refs.~\cite{KM,KMR2}.
First note that there are only two
independent color structures for this subprocess. The $s$-channel matrix
element is a sum of two parts, each of which is proportional to one of the two
independent color structures.
We combine terms with the same color structures
of the three (e.g. $s$, $t$, and $u$) production channels. Finally, we remove
the heavy-antiquark momentum $p_4$ using energy-momentum conservation
and use on-shell conditions for the gluons ($p_1\cdot\epsilon_1=0$ and 
$p_2\cdot\epsilon_2=0)$ and the heavy quark 
($\bar{u}_{3}{\rm \hspace{-.04in}}\not p_3= \bar{u}_{3}m$). We then obtain
the two color-linked LO matrix elements
\be
\label{lomatrix}
M_{\rm LO,t}=i T^b T^a \hat{M}/t,
\qquad
M_{\rm LO,u}=i T^a T^b \hat{M}/u,
\ee
with
\be
\label{mbrn}
s \hat{M}=  \gamma^{\mu} {\rm \hspace{-.07in}}\not p_1\gamma^{\nu} s
+ 2 \gamma^{\mu} p_1^{\nu} t
- 2 \gamma^{\nu} p_2^{\mu} t - 2 \gamma^{\nu} p_3^{\mu} s -
2 {\rm \hspace{-.07in}}\not p_1 g^{\mu\nu} t  .
\ee
It can be verified that the function $\hat{M}$ is $t \lra u$ symmetric, and
consequently the color-linked Born amplitudes $M_{\rm LO,t}$ and $M_{\rm LO,u}$
turn into one another under $t \leftrightarrow u$.

We then square the full Born matrix element $M_{\rm LO,t}+M_{\rm LO,u}$ and do the
spin and color sums to obtain
the LO amplitude,
\be
\label{lo}
|M|_{\rm LO}^2 = \frac{d_A}{2} 
                               \left(  C_F \frac{s^2}{t u} - N_C
                        \right)  |\hat{M}|^{2} \equiv B,
\ee
where we have factored out a color-reduced Born term $|\hat{M}|^{2}$, which reads
\ba
\label{brn}
\nn
|\hat{M}|^2 = 8 \Big\{
            \frac{t^2 + u^2}{s^2} + 4 \frac{m^2}{s} - 4 \frac{m^4}{t u} && \\
              - \ep\, 2 (1 - \frac{t u}{s^2}) + \ep^2
\Big\} \equiv \hat{B}  . &&
\ea

The expression in Eq.~(\ref{lo}) for the LO amplitude agrees with the 
well-known result in $n$ dimensions (see e.g. Ref.~\cite{Been}). 
Note that, by using the prescription of Ref.~\cite{Slaven}, we were
able to avoid the introduction of ghost contributions which would otherwise
arise from the square of the right-most three-gluon coupling
amplitude in Fig.~\ref{fig:born}. In our case the prescription of 
Ref.~\cite{Slaven}
consists in the use of on-shell conditions for external gluons, i.e.
$p_1\cdot\epsilon_1=0$ and $p_2\cdot\epsilon_2=0$, and the exclusion
of the heavy-antiquark momentum via $p_{4}=p_1+p_{2}-p_{3}$.
When squaring amplitudes, we sum over the two helicities of the gluons using
the Feynman gauge, i.e. we use
\begin{equation}
\sum_{\lambda= \pm 1} \epsilon^{\mu}(\lambda)\epsilon^{\nu}(\lambda)= - g^{\mu \nu}\,.
\end{equation}
The use of the framework set up in Ref.~\cite{Slaven} has the advantage in the
non-Abelian case that one can omit ghost contributions when squaring the
amplitudes. Using the above
on-shell conditions already at the amplitude level means that one takes
full advantage of the gauge invariance of the problem when squaring the amplitudes.
Thus, in general, the results for the different channels will not be identical
to the ones which would be obtained using 't Hooft-Feynman gauge throughout.

Folding the one-loop matrix elements (see Figs.~\ref{fig:ggnlot} and
\ref{fig:ggnlos}) with the
LO Born term (see Fig.~\ref{fig:born}), 
one obtains the virtual part of the NLO result.

As concerns the one-loop matrix elements, we shall use the
one-loop matrix elements of Refs.~\cite{KM,KMR2}
to compute the virtual NLO contribution up to ${\mathcal O}(\ep^2)$ in terms of
the coefficient functions (\ref{Dexp}) of the scalar master integrals. 
However, in Ref.~\cite{KM}, where expressions for the
NLO matrix elements up to ${\mathcal O}(\ep^0)$ are given,
the values for the scalar coefficient functions in terms of 
logarithms and dilogarithms are
substituted directly. Therefore, we had to recalculate
the corresponding expressions from Ref.~\cite{KM} for the matrix elements in order
to have a uniform result in terms of scalar coefficient functions. 
This has allowed 
us to retrieve and use relations between coefficients of the scalar
coefficient functions in the result for different 
orders of the Laurent series expansion in
$\ep$. We will comment on these relations later on. 

We also mention that we had to
regroup and rearrange various terms in the one-loop amplitudes from 
Refs.~\cite{KM,KMR2} according to 
the three independent color structures
in order to bring the pole terms into agreement with the form suggested in 
Ref.~\cite{Catani}.
In the gluon fusion case treated here, there are three independent color structures
in the one-loop amplitudes,
e.g. $T^b T^a, T^a T^b$, and $\delta^{ab}$.
As in the LO case, one also has to exclude the heavy-antiquark momentum $p_4$
from the one-loop amplitude expressions.
As a result of the above two steps, the pole terms of our new matrix elements
became proportional to the LO color-linked amplitudes (\ref{lomatrix}).
In all our subsequent calculations, we shall use only these matrix elements.

The NLO virtual corrections to heavy-flavor hadroproduction have been
calculated before for the $g g\to Q{\bar Q}$ case.
Nevertheless, one cannot find explicit
separate results for the virtual corrections in the literature although 
Ref.~\cite{Been} provides analytic results for the combined 
``virtual+soft'' contributions. 
We have therefore recalculated 
the virtual NLO contribution to $gg$ fusion. In fact, we have
calculated the virtual NLO results up to ${\mathcal O}(\ep^2)$. As it turns
out, use of
the expressions for the NLO virtual ${\mathcal O}(\ep^1)$- and
${\mathcal O}(\ep^2)$-contributions
considerably simplify the presentation of the corresponding NNLO results in as
much as they appear as important building blocks in the NNLO results.

Next we fold the pole, 
finite, $\mathcal O (\ep^1)$ and $\mathcal O (\ep^2)$ terms of our NLO
matrix element with the LO matrix element. In dimensional regularization, 
the trace evaluation in $n=4-2\ep$ dimensions will lead to
terms of order ${\mathcal O}(\ep^1)$ and ${\mathcal O}(\ep^2)$ 
when multiplied with the pole and finite
terms, as well as to the terms of 
${\mathcal O}(\ep^3)$ and ${\mathcal O}(\ep^4)$ when multiplied 
with the  $\mathcal O (\ep^1)$ and $\mathcal O (\ep^2)$ terms of the
squared amplitude, 
respectively. In the following we will disregard terms of 
${\mathcal O}(\ep^3)$ and ${\mathcal O}(\ep^4)$ as they do not
contribute to the finite part of the NNLO result.

Before presenting our result for the NLO matrix element, we would like to
comment on its color structure.
We have decomposed our matrix elements according
to the
following three independent color structures:
\ba
\label{colnlo}
\delta^{ab} \,\, {\rm Tr} (T^a T^b) &\!=\!& \frac{d_A}{2} ,    \\
\nn
{\rm Tr} (T^b T^a) \,\, {\rm Tr} (T^b T^a)  &\!=\!&
  \frac{d_A}{2} C_F, \\
\nn
{\rm Tr} (T^b T^a) \,\,{\rm Tr} (T^a T^b) &\!=\!&
                  \frac{d_A}{2} (C_F - \frac{N_C}{2}) \, .
\ea

At NLO, the final spin and color summed matrix element can be written as a
sum of five terms:
\ba
\label{nlogg}
\nn
|M|^2_{{\rm Loop}\times{\rm Born}} =
                  g_s^2 \sqrt\mathcal C \,\,{\rm Re}\Big[   \frac{1}{\ep^2} W^{(-2)}(\ep) +
                      \frac{1}{\ep} W^{(-1)}(\ep)   &&     \\ \nn
+ W^{(0)}(\ep) + \ep W^{(1)}(\ep) + \ep^2 W^{(2)}(\ep) \Big],   &&   \\
\ea
where $\mathcal C$ has been defined in Eq.~(\ref{common}). The notation
$|M|^2_{{\rm Loop}\times{\rm Born}}$ means that one is retaining only the
${\mathcal O} (\alpha_{s}^3)$ part of $|M|^{2}$.

The first two coefficient functions in Eq.~(\ref{nlogg}) have a rather simple
structure:
\ba
\label{polesnlo}
W^{(-2)}(\ep) &\!\!=\!\!& - 4 N_C B \, ,    \\
\nn
W^{(-1)}(\ep) &\!\!=\!\!&  d_A \hat{B} \Big( \frac{s^2}{t u} f_{\delta}
         + (C_F - \frac{N_C}{2}) (f_{t} + f_{u})     \\
\nn   &&
+ C_F \frac{u}{t} f_{t} + C_F \frac{t}{u} f_{u}
\Big) \, ,
\ea
where $B$ and $\hat{B}$ are the LO terms defined in Eqs.~(\ref{lo}) and
(\ref{brn}). We have also introduced new functions,
\ba
\label{flogs}
\nn
f_{\delta} &\!\!=\!\!&  \frac{1}{2} \ln\frac{s}{m^2} + \frac{t}{s} \ln\frac{-t}{m^2}
                     + \frac{u}{s} \ln\frac{-u}{m^2}
                     + \frac{2 m^2 - s}{2s\beta} \ln x,   \\
\nn
f_{t} &\!\!=\!\!&   \nc \ln\frac{s}{m^2} + 2 \nc \ln\frac{-t}{m^2}
                       - 2\cf - \beta_0     \\
\nn   &&
+ (2\cf - \nc) \frac{2 m^2 - s}{s\beta} \ln x ,    \\
f_{u} &\!\!=\!\!&  f_{t}|_{t\lra u},
\ea
where $\beta = \sqrt{1-4m^2/s}$ is the heavy-quark velocity and 
$\beta_{0}$ is defined after Eq.~(\ref{renormconstants}).

One should keep in mind that the overall Born term factors $B$ and $\hat{B}$
contain terms multiplied by $\ep$ and $\ep^2$. Therefore, if the expressions 
for $B$ and $\hat{B}$, given in 
Eqs.~(\ref{lo}) and (\ref{brn}), are substituted
in $W^{(-2)}$ and $W^{(-1)}$, we will obtain additional
${\mathcal O}(\ep^{-1})$ and finite terms  from the first two terms of
Eq.~(\ref{nlogg}).

The third term in Eq.~(\ref{nlogg}) reads
\be
\label{finNLO}
W^{(0)}(\ep) \equiv  F_{\rm NLO}^{(0)} \,  ,
\ee
where we have constructed the following generic functions:
\be
\label{nloterm}
F_{\rm NLO}^{(j)} = {\cal W}_1^{(j)} + {\cal W}_2^{(j)} \, ,
\ee
with
\ba
\label{nloterms}
\nn
{\cal W}_1^{(j)} &\!\!=\!\!&   -\frac{d_A}{2}\Big[
                   \frac{s}{tu} F_1^{(j)}  + \Big\{
                   \frac{1}{u} \Big( \frac{s}{t} \cf + \frac{\nc}{2} \Big)
                        ( F_2^{(j)} + F_3^{(j)} )               \\
\nn &&
                                  + \,\, ( t\lra u ) \Big\} \Big] ,   \\
{\cal W}_2^{(j)} &\!\!=\!\!&  
-\frac{2 B \beta_0}{(1+j) !} \ln^{1+j}\frac{m^2}{\mu^2} \, .
\ea
The three functions $F_1, F_2$, and $F_3$ are defined as follows:
\ba  
\label{F1}  \nn
F_1^{(j)} \!\!&=&\!\! \sum_I (a_I + \ep a_I^{(\ep)} + \ep^2 a_I^{(\ep^{2})}) I^{(j)} ,  
\\
\nn
{\rm with} && I^{(j)}=\{B_2,B_5,C_1,C_2,C_{2u},C_3,C_{3u},C_4,C_5,C_6,    \\
\nn  &&
      \qquad \qquad            D_1,D_{1u},D_2,D_{2u},D_3\}^{(j)};    \\[2mm]
\label{F2}
F_2^{(j)} \!\!&=&\!\! \sum_I (b_I + \ep b_I^{(\ep)}) I^{(j)} ,   \\
\nn
{\rm with} && I^{(j)}=\{1,B_2,B_5,C_1,C_4,C_5,C_6\}^{(j)};        \\[2mm]
\label{F3} \nn
F_3^{(j)} \!\!&=&\!\! \sum_I (c_I + \ep c_I^{(\ep)} + \ep^2 c_I^{(\ep^{2})}) I^{(j)},    
\\
\nn 
{\rm with} && I^{(j)}=\{1,B_1,B_2,B_5,C_1,C_2,C_3,C_4,C_5,C_6,           \\
\nn  &&
           \qquad \qquad                 D_1,D_2 \}^{(j)} \, .
\ea

For $I=1$ one has $I^{(j)}\equiv 1$, otherwise
$I^{(j)}\equiv B_1^{(j)}, C_2^{(j)}$ etc. 
In other words, the summation index $I$
runs over the scalar integral coefficient functions, while the
coefficient functions $a_I, a_I^{(\ep)}, a_I^{(\ep^{2})}$ etc. denote the
explicit dependence on $s,t$ and $m^2$.
These coefficient functions are presented in Appendix~A.
Note that index $j$ takes the same value for all the coefficient functions 
in Eq.~(\ref{F2}) as well as in similar equations that will follow.

The additional subscript ``u'' in some of the scalar coefficient functions in
the expression for $F_1^{(j)}$ (such as $C_{2u}^{(j)}$) is to be understood  
as an operational definition
prescribing a $(t\lra u)$ interchange in the argument of that function, i.e.
$C_{2u}^{(0)}=C_{2}^{(0)}\big|_{t\lra u}$ etc.

Note that ${\cal W}_2^{(j)}$ is only contributed to by the renormalization 
procedure.
Of course, all the remaining ${\mathcal O} (\ep)$ terms 
(e.g. $W^{(1)}(\ep)$ and $W^{(2)}(\ep)$, as well as those coming from 
$W^{(-1)}(\ep)/\ep$ and $W^{(0)}(\ep)$) 
should be
disregarded in the NLO final result in Eq.~(\ref{nlogg}). 
It is important to note
that $F_{\rm NLO}^{(0)}|_{\ep=0}$ is not formally the full finite part of the
NLO result in dimensional regularization, 
but it results from folding the finite part of our original NLO matrix element
with the LO one. Another part of the finite result comes from the first two terms
in Eq.~(\ref{nlogg}), as mentioned before Eq.~(\ref{finNLO}).
However, one should realize that the first two terms in Eq.~(\ref{nlogg}) would
be cancelled with the corresponding parts from the real bremsstrahlung diagrams.
Given the overall factor, Eq.~(\ref{ceps}), the term $F_{\rm NLO}^{(0)}$ evaluated 
for ${\ep=0}$ represents the finite part of the virtual one-loop NLO result.

Our ${\mathcal O} (\ep^{-2})$, ${\mathcal O} (\ep^{-1})$, and ${\mathcal O} (\ep^{0})$
NLO results in Eq.~(\ref{nlogg})
were analytically compared with the corresponding
results obtained
in Ref.~\cite{Ellis}, which were kindly provided to us in a Schoonschip
format by the 
authors \cite{EllisNason}. We obtained complete agreement.

The fourth term in Eq.~(\ref{nlogg}) is a result of folding the 
${\mathcal O} (\ep)$ term of the matrix element with the Born term. 
Because of the $n$-dimensional traces, one also obtains terms of 
${\mathcal O} (\ep^2)$ and ${\mathcal O} (\ep^3)$. As mentioned before, 
we will only retain terms of ${\mathcal O}(\ep)$ and ${\mathcal O}(\ep^2)$.
We have
\be
\label{epterm}
W^{(1)}(\ep) = F_{\rm NLO}^{(1)} + F_{\rm NLO, \ep}^{(0)},
\ee
where
\ba  \nn
F_{\rm NLO, \ep}^{(j)}  &\!\!=\!\!&    d_A \Big[ F_4^{(j)} - 
                         \Big\{ \Big( \cf + \frac{\nc}{2}\frac{t}{s}\Big)
                   (F_5^{(j)} + \cf F_6^{(j)} \\
&&
+ \nc F_7^{(j)}) + (t\lra u) \Big\} \Big] \, .
\ea
Here
\ba
\label{F4}   \nn
F_4^{(j)} &\!\!=\!\!& \sum_I (d_I^{(\ep)} + \ep d_I^{(\ep^2)}) I^{(j)} ,  \\
\nn
{\rm with} && I^{(j)}=\{B_2,B_5,C_1,C_2,C_{2u},C_3,C_{3u},C_4,C_5,C_6,        \\
\nn  &&
      \qquad \qquad                D_1,D_{1u},D_2,D_{2u},D_3\}^{(j)};    \\[2mm]
\label{F5}  \nn
F_5^{(j)} &\!\!=\!\!&  \sum_I (e_I^{(\ep)} + \ep e_I^{(\ep^2)}) I^{(j)} ,   \\
\nn
{\rm with} && I^{(j)}=\{1,B_2,B_5,C_5\}^{(j)};        \\[2mm]
\label{F6}
F_6^{(j)} &\!\!=\!\!& \sum_I (g_I^{(\ep)} + \ep g_I^{(\ep^2)} ) I^{(j)},    \\
\nn
{\rm with} && I^{(j)}=\{1,B_1,B_2,C_2,C_5,C_6,D_1 \}^{(j)};     \\[2mm]
\label{F7}  \nn
F_7^{(j)} &\!\!=\!\!& \sum_I (h_I^{(\ep)} + \ep h_I^{(\ep^2)} ) I^{(j)} ,   \\
\nn
{\rm with} && I^{(j)}=\{1,B_1,B_2,B_5,C_1,C_2,C_3,C_4,C_5,C_6,   \\
\nn  &&
      \qquad \qquad                  D_1,D_2\}^{(j)} \, .
\ea
The coefficients $d_I, e_I, g_I, h_I$ are presented in Appendix~B.
Note that the first term in Eq.~(\ref{epterm}) in nothing but the NLO term of
Eq.~(\ref{finNLO})
with indices of the
coefficient functions of the scalar master integrals and the power of
the logarithm that multiplies $\beta_0$ shifted upwards by one.

The last term in Eq.~(\ref{nlogg}) is a result of folding the
${\mathcal O} (\ep^2)$ term of the matrix element with the Born term.
Because of the $n$-dimensional traces, one also obtains terms of 
${\mathcal O} (\ep^3)$ and ${\mathcal O} (\ep^4)$, which are omitted as before. 
For the ${\mathcal O}(\ep^2)$ terms we obtain
\be
\label{ep2term}
W^{(2)}(\ep) = F_{\rm NLO}^{(2)} + F_{\rm NLO, \ep}^{(1)} + 
               F_{\rm NLO, \ep^2}^{(0)} \; ,
\ee
where
\ba  \nn
F_{\rm NLO, \ep^2}^{(j)}  &\!\!=\!\!&    d_A \Big[ F_8^{(j)} -
                         \Big\{ \Big( \cf + \frac{\nc}{2}\frac{t}{s}\Big)     
                   (F_9^{(j)} + \cf F_{10}^{(j)} \\
&&
+ \nc F_{11}^{(j)}) + (t\lra u) \Big\} \Big] \, .
\ea
Here
\ba
\label{F8}  \nn
F_8^{(j)} &\!\!=\!\!& \sum_I k_I^{(\ep^2)} I^{(j)} ,  \\
\nn 
{\rm with} && I^{(j)}=\{C_1,C_2,C_{2u},C_3,C_{3u},C_4,C_5,C_6,D_1,D_{1u},     \\
\nn  &&
      \qquad \qquad                    D_2,D_{2u},D_3\}^{(j)};    \\[2mm]
\label{F9}   \nn
F_9^{(j)} &\!\!=\!\!& \sum_I l_I^{(\ep^2)} I^{(j)} ,   \\
\nn
{\rm with} && I^{(j)}=\{1,B_2,B_5,C_5\}^{(j)};        \\[2mm]
\label{F10}
F_{10}^{(j)} &\!\!=\!\!& \sum_I  m_I^{(\ep^2)} I^{(j)},    \\
\nn 
{\rm with} && I^{(j)}=\{1,C_2,C_5,C_6,D_1 \}^{(j)};        \\[2mm]
\label{F11}  \nn
F_{11}^{(j)} &\!\!=\!\!& \sum_I n_I^{(\ep^2)} I^{(j)} ,   \\
\nn 
{\rm with} && I^{(j)}=\{1,B_5,C_1,C_2,C_3,C_4,C_5,C_6,D_1,D_2\}^{(j)} \, .
\ea
The coefficients $k_I, l_I, m_I, n_I$ are presented in Appendix~C.
We mention that the functions $F_1, F_4$, and $F_8$ are $(t\lra u)$
symmetric.



\section{\label{singular}
SINGULARITY STRUCTURE OF THE NNLO SQUARED AMPLITUDE
}


The NNLO final spin and color summed squared matrix element can be
written down as a sum of five terms:

\begin{eqnarray}
\label{nnlo}
\frac{1}{\mathcal C} |M|^2_{{\rm Loop}\times{\rm Loop}} &\!\!=\!\!&
                  {\rm Re}\Big[    \frac{1}{\ep^4} V^{(-4)}(\ep) +
                      \frac{1}{\ep^3} V^{(-3)}(\ep) \\
& &\mbox{\hspn}                    +\frac{1}{\ep^2} V^{(-2)}(\ep)
                     + \frac{1}{\ep} V^{(-1)}(\ep) +
                                    V^{(0)}(\ep) \Big] ,
\nonumber
\end{eqnarray}
where $\mathcal C$ has been defined in Eq.~(\ref{common}). 
Note that Eq.~(\ref{nnlo}) is
{\it not} a Laurent series expansion in $\ep$ since the coefficient functions
$V^{(m)}(\ep)$ are functions of $\ep$ as explicitly annotated in Eq.~(\ref{nnlo}).
It is nevertheless useful to write the NNLO one-loop squared result in the
form of Eq.~(\ref{nnlo}) in order to exhibit the explicit $\ep$ structures.
All five coefficient functions $V^{(m)}(\ep)$ 
are bilinear forms in the coefficient functions that define the
Laurent series expansion of the scalar master integrals (\ref{Dexp}).
Some of these coefficient
functions are zero and some of them are just numbers or simple logarithms.
In the latter case, we have 
substituted these numbers or logarithms for the coefficient functions
$V^{(m)}$ in the five terms above. This has been done for all the 
scalar coefficient functions that multiply poles, i.e. for scalar functions 
with negative subscripts $I^{(-2)}$ and $I^{(-1)}$,
as well as for the whole scalar functions $A^{(i)}$, $B_3^{(i)}$, and $B_4^{(i)}$.
%

%
%
We found that a significant part of the NNLO results 
can be expressed in terms of the $\ep$ expansion of the NLO contribution.
In particular, we will need the NLO expansion up to $\ep^2$.
Therefore, in this section, we will make full use of the results derived in 
Sec.~\ref{nlo}.

Before proceeding further, we note that there are no additional color
structures appearing in 
the NNLO calculation for $gg$ fusion in addition to the ones already 
presented in Eq.~(\ref{colnlo}):
they are just linear combinations of the ones in the NLO case. 
This is in contrast to 
the $q\bar{q}$ subprocess, where the NNLO color structures exhibit 
much higher complexity and richness \cite{KMRqq} relative to the 
NLO ones.

The two most singular terms in Eq.~(\ref{nnlo}) are proportional
to the Born $B$ and color-reduced Born $\hat{B}$ terms defined in 
Eqs.~(\ref{lo}) and (\ref{brn}), respectively. One has
\ba
\label{eps43}
V^{(-4)}(\ep) &\!=\!& 4 N_C^2 B ,    \\
\nn
V^{(-3)}(\ep) &\!=\!&  - 2 N_C   W^{(-1)}(\ep) \, ,
\ea
where $W^{(-1)}(\ep)$ is given in Eq.~(\ref{polesnlo}) and is nothing but 
the full coefficient of the single-pole NLO result.

For the $1/\ep^2$ term we obtain
\ba
\label{eps2} \nn
V^{(-2)}(\ep) &\!=\!& d_A \hat{B} \Big[
     \frac{s^2}{t u} |f_{\delta}|^2 
   + \frac{1}{2} C_F \Big( \frac{u}{t} |f_{t}|^2 + \frac{t}{u} |f_{u}|^2 \Big)
\\ \nn  &&
         - s f_{\delta}^\ast \Big( \frac{1}{t} f_t 
                                                     + \frac{1}{u} f_u  \Big)
             + (C_F - \frac{N_C}{2}) f_t^\ast f_u
              \Big]             \\
&&
- 2 \nc F_{\rm NLO}^{(0)} \, ,
\ea
where the functions $f_{\delta}, f_t$, and $f_u$ 
above are the same as those in
Eq.~(\ref{flogs}), but now with the imaginary parts retained, i.e. one has
the following replacements:
\be
\label{imaglns}
\ln\frac{s}{m^2} \ra \ln\frac{s}{m^2} - i\pi, \qquad
\ln x \ra \ln x + i\pi.
\ee
This reflects the fact that, contrary to the NLO calculation, one has to
keep the imaginary parts in the NNLO calculation 
as emphasized in the Introduction. It should be clear
that the completion (\ref{imaglns}) has to be done everywhere in the NNLO
calculation whenever the logarithms (\ref{imaglns}) appear in bilinear
forms multiplying complex functions.

The last term $- 2 \nc F_{\rm NLO}^{(0)}$ in Eq.~(\ref{eps2}) is obtained from 
folding the
${\mathcal O}(\ep^{-2})$ singular term of the matrix element with its finite 
part,
while the remaining parts result from folding the single poles.
Note that when one substitutes the Laurent expansions for $\hat{B}$ 
and $F_{\rm NLO}^{(0)}$,
one gets additional $1/\ep$ poles and finite terms in Eq.~(\ref{eps2}).

The structure of the fourth term in Eq.~(\ref{nnlo}) is somewhat 
more complicated. 
One has
\ba
\label{eps}
\nn
V^{(-1)}(\ep) &\!\!=\!\!&  \frac{\beta_0}{2 N_C} \ln(\frac{m^2}{\mu^2}) V^{(-3)}(\ep)
             + {\cal S}_{1}^{(0)} - 2 N_C W^{(1)} (\ep) ,  \\
\ea
where we have introduced new functions
\ba
\nn
{\cal S}_{1}^{(j)} &\!\!=\!\!& - \frac{d_A}{4} \frac{s}{tu}
\Big( L_1^\ast F_1^{(j)} + L_2^\ast F_2^{(j)} +  L_2^\ast F_3^{(j)}
              + (t\lra u)  \Big) ,   \\
\ea
with
\ba
L_1 &\!\!=\!\!& 2 f_\delta - \frac{u}{s} f_t - \frac{t}{s} f_u  \; , \\ \nn 
L_2 &\!\!=\!\!& 2 f_\delta - 2 \cf \frac{u}{s} f_t - (2\cf - \nc) \frac{t}{s} f_u 
\; .
\ea
%
%
The first two terms in Eq.~(\ref{eps}) arise from 
folding the single-pole terms in the original matrix element
with its finite $\mathcal O(\ep^0)$ part. The last term is due
to the interference of $\mathcal O(\ep^{-2})\times \mathcal O(\ep)$ terms in
the original matrix element. This pole term is due to the Laurent series
expansion of the original matrix element and cannot be
deduced from the knowledge of the NLO terms alone.
The function $W^{(1)}(\ep)$ is defined in Eq.~(\ref{epterm}), 
while the functions $F_1^{(j)}, F_2^{(j)}$,
and $F_3^{(j)}$ are given by Eq.~(\ref{F1}).

When one substitutes the Laurent expansions for $F_1^{(0)}$, 
$F_2^{(0)}$, $F_3^{(0)}$, and $W^{(1)}(\ep)$, 
one gets finite and ${\mathcal O}(\ep)$ terms in Eq.~(\ref{eps}).
However, since
we are only interested in the Laurent series expansion up to the finite
term, these $\mathcal O(\ep)$ contributions can be omitted as before.



\section{\label{finite}
STRUCTURE OF THE FINITE PART
}

In this section, we present the finite part of our result. 
In the course of our calculation, we have made full use 
of the results presented in Sec.~\ref{nlo}, e.g. of our detailed study of the 
NLO structure of the Laurent series expansion up to $O(\ep^2)$. 
As a consequence, we can present a large part of our results 
for the finite part in a 
surprisingly concise and closed form. 
We decompose the finite part into several pieces, as
\be
\label{fin}
V^{(0)}(\ep) = {\rm Re}\left[V_{11}^{(0)} + V_{22}^{(0)}+ V_{00}^{(0)}\right].
\ee

The first two terms originate from the interference of the
$\mathcal O(\ep^{-1})\times \mathcal O(\ep)$ and
$\mathcal O(\ep^{-2})\times \mathcal O(\ep^2)$ pieces of the initial
matrix element, respectively.
Each of them can be conveniently presented in a very compact form:
\ba
\nn
V_{11}^{(0)} &\!\!=\!\!& \frac{d_A}{2} \hat{B} \beta_0 \ln^2(\frac{m^2}{\mu^2}) 
       \Big[ - \frac{s^2}{t u} f_{\delta} 
             + \Big(\frac{s}{t} C_F + \frac{N_C}{2}\Big) f_t          \\ 
\nn &&
             + \Big(\frac{s}{u} C_F + \frac{N_C}{2}\Big) f_u \Big]    \\
&&
+ {\cal S}_{1}^{(1)}  +  {\cal S}_{2}^{(0)} ,
\ea
where we have introduced one more function,
\ba
\nn
{\cal S}_{2}^{(j)} &\!\!=\!\!& 
                                   d_A \Big[ L_1^\ast F_4^{(j)} - 
                         \Big\{ \frac{L_2^\ast}{2} 
(F_5^{(j)} + \cf F_6^{(j)} + \nc F_7^{(j)})    \\
            &&  + (t\lra u) \Big\} \Big] \, ,
\ea
Similarly, for the second term in Eq.~(\ref{fin}), we write
\be
V_{22}^{(0)} = -2 N_C W^{(2)}(\ep)  ,
\ee
with $W^{(2)}(\ep)$ defined in Eq.~(\ref{ep2term}). 
Note again that the $\mathcal O(\ep)$ and $\mathcal O(\ep^2)$
terms in the above expressions
for $V_{11}^{(0)}$ and $V_{22}^{(0)}$ can be disregarded.
We mention that
the scalar coefficient functions with the superscript ``2'' above involve
multiple polylogarithms of weight and depth 4.

We emphasize that the quasifactorized forms of all the expressions given in
this paper hold only when one retains the full $\ep$ dependence in the Born
and NLO terms.

The last term in Eq.~(\ref{fin}) comes from the square of the $O(\ep^0)$
term of the matrix element, which
can be written as 
\be
\label{ep0ep0}
V_{00}^{(0)} = -\beta_0 \ln(\frac{m^2}{\mu^2}) 
      \Big[ F_{\rm NLO}^{(0)} - \frac{1}{2} {\mathcal W}_2^{(0)} \Big]
        + Y \, ,
\ee
where $F_{\rm NLO}^{(0)}$ and $\mathcal W_2^{(0)}$ are given in 
Eqs.~(\ref{nloterm}) and (\ref{nloterms}). 
We found that the last term $Y$ in Eq.~(\ref{ep0ep0}) also possesses the 
quasifactorization
properties discovered in a recent paper \cite{KMRqq}. For instance, the result
can also be 
written down as a sum of bilinear products, where each of the factors are
linear combinations of scalar integral coefficient functions multiplied
by some combinations of kinematic variables. 
However, because of the great number of Laurent structures appearing 
in the original matrix element for the $gg$ fusion subprocess, the length
of the final expressions does not allow us to present the results in this 
paper. Also, we were not able to find the optimal way to organize the different
contributions in $Y$ as in Ref.~\cite{KMRqq}, as not all the powers of 
common numerators and denominators
cancel out. Therefore, we have opted to supply the results on the finite term 
$Y$ in a separate electronic file.

In the finite contribution of Eq.~(\ref{fin}), one notices the interplay of the
product of powers of $\ep$ resulting from the Laurent series expansion of the
scalar integrals [cf. Eq.~(\ref{Dexp})] on the one hand and powers of $\ep$
resulting from doing the spin algebra in dimensional regularization on the other
hand. For example, for the finite part
one has a contribution from $C_6^{(-1)}B_1^{(0)\ast}$ as well as a contribution from
$C_6^{(-1)}B_1^{(1)\ast}$. Terms of the type $C_6^{(-1)}B_1^{(0)\ast}$, where the
superscripts corresponding to
$\ep$ powers do not compensate, would be absent in regularization schemes
where traces are effectively taken in four dimensions, i.e. in the so-called
four-dimensional schemes or in dimensional reduction (DRED).

We emphasize that all our factorized
results given in this paper [except for the expression for $Y$ 
in Eq.~(\ref{ep0ep0})] take up about
22 Kb of hard disk space. This has to be compared with the length of the
original, untreated FORM output.
The original computer output for the corresponding one-loop squared
cross section of the $g g\ra Q{\overline Q}$ subprocess
turned out to be very long and took up about 85~MB of hard disk space.
Therefore, the reduction is of the order of $10^3$--$10^4$ in the present case.

As a final remark we want to emphasize that we have done two independent
calculations using REDUCE \cite{reduce} and FORM \cite{form} when squaring
the one-loop amplitudes. 
The results of both calculations agree. 
Casting the results into the compact forms presented in this paper 
was done with the help of the REDUCE Computer Algebra System.


\section{\label{summary}
CONCLUSIONS
}


We have presented analytical ${\cal O}(\alpha_s^4)$ NNLO results
for the one-loop squared contributions to heavy-quark pair production in the
gluon-gluon fusion reaction. 
The corresponding result for photon-photon
fusion has already been presented in Ref.~\cite{gamgam}, while results for the
photon-gluon fusion process can be obtained from Ref.~\cite{KMR2} after some 
color factor adjustments. 
As concerns hadroproduction of heavy quarks, 
the results of the present paper, together with a recent
publication on $q\bar{q}$ production \cite{KMRqq}, complete the derivation of 
the one-loop squared contributions to the
hadroproduction of heavy quarks at NNLO with the heavy-quark mass dependence
fully retained. Our results form part of the NNLO description of heavy-quark
pair production relevant for the NNLO analysis of ongoing experiments at the
TEVATRON and the LHC.

A large part of our analytical results are presented in a very compact form. 
The singular contributions proportional to $\ep^{-4}, \ep^{-3}$, and $\ep^{-2}$ 
are entirely given in terms of LO and NLO contributions, 
whereas the $\ep^{-1}$ contributions contain some true NNLO
structure in addition to LO and NLO structures.
Since the LO and NLO terms are themselves expanded in Laurent series, 
this implies that our singular contributions are not true 
(in a mathematical sense) Laurent series in $\ep$. 
We believe that our representation of the singular contributions 
has structural advantages in as much as it will be simpler to match our
singular structures onto the singular structures of the other classes of 
contributions. Also, our representation is convenient if one wants 
to convert our expressions to different regularization schemes such as DRED
(see e.g. Ref.~\cite{SignerStock}).
If needed, our singular contributions can easily be converted into true
Laurent series expansions since our expressions are very compact.

Because of our representation of the singular parts, we obtained quasifactorized 
expressions for a large part of the finite contributions. 
Writing our analytical results in factorized forms
led to a reduction of the length of the original output by a 
factor of $10^3$--$10^4$, which will lead to a dramatic reduction of the CPU 
time needed in numerical evaluations. 


The present paper deals with unpolarized gluons in the initial state and
unpolarized heavy quarks in the final state.
Since our results for the original matrix elements contain the full spin
information of the process, an extension to the polarized case with
polarization in the initial state and/or in the final state including spin
correlations would be possible.


Analytical results in electronic format for the coefficients given in 
the Appendices 
as well as for the term $Y$ in Eq.~(\ref{ep0ep0}) are readily available
\cite{offering}.

\begin{acknowledgments}
We would like to thank J.~Gegelia, A.~Kotikov, G.~Kramer, and O.~Veretin  
for useful discussions. 
We are very grateful to R.K.~Ellis and P.~Nason for swift response and 
for providing the 
electronic files of their analytical one-loop virtual NLO results. 
We also acknowledge helpful  
communications with W.~Beenakker, 
I.~Bojak, I.~Schienbein, J.~Smith, and H.~Spiesberger. 
Z.M. would like to thank the Particle Theory group of the
Institut f{\"u}r Physik, Universit{\"a}t Mainz for hospitality, where this
work has started. 
The work of Z.M. was supported in part 
by the German Research Foundation DFG 
through Grants No.~KN~365/7-1 and No.~KO~1069/11-1, and 
by the Georgia National Science 
Foundation through Grant No.~GNSF/ST07/4-196.
M.R. was supported by the Helmholtz Gemeinschaft HGF 
under Contract No. VH-NG-105.

{\it Note added}.-- While finalizing our manuscript for publication,
we became aware of the preprint \cite{Aybat}
by Anastasiou and Mert Aybat, who also discuss the NNLO one-loop
squared gluon fusion production of heavy-quark pairs.
\end{acknowledgments}

\appendix
\section{}

First, we write down a few abbreviations that we use throughout the
paper:
\ba   
\label{a1}
\nn   &
\beta=\sqrt{1-4m^2/s}, \qquad   D=m^2 s - t u, & \\
&
z_2=s + 2 t, \qquad     z_{2u}=s + 2 u,  & \\
\nn   &  
z_t=2 m^2 + t,       \qquad    z_u=2 m^2 + u.  &
\ea
Note that $D$ in Eq.~(\ref{a1}) is {\it not} the space-time dimension.

Here we present the expressions for all the coefficients $a_I, b_I, c_I$ 
appearing in Eq.~(\ref{F2}):

\begin{eqnarray}
\label{eq:b12}
  a_{B_2} &\! =\! &
  16 D/(s \beta^2)
\nonumber
\; ,
\quad
\\[2mm]
\label{eq:b15}
\nonumber
  a_{B_5} &\! =\! &
               -a_{B_2}
\; ,
\quad
\\[2mm]
\label{eq:c11}
\nonumber
  a_{C_1} &\! =\! &
4 ( 8 m^4 - z_2^2/s (2 m^2 - s + 2 m^2/\beta^2) )
\; ,
\quad
\\[2mm]
\label{eq:c12}
\nonumber
  a_{C_2} &\! =\! &
8 t/s (4 m^2 z_t + 2 s t + t^2)
\; ,
\quad  
\\[2mm]
\label{eq:c12u}
\nonumber
  a_{C_{2u}} &\! =\! &
    a_{C_2} (t\lra u)
\; ,
\quad  
\\[2mm]
\label{eq:c13}
\nonumber
  a_{C_3} &\! =\! &
8 t/s (4 m^2 z_t + t z_2)
\; ,
\quad  
\\[2mm]
\label{eq:c13u}
\nonumber
  a_{C_{3u}} &\! =\! &
   a_{C_3} (t\lra u)
\; ,
\quad  
\\[2mm]
\label{eq:c14}
\nonumber
  a_{C_4} &\! =\! &
4 (4 m^2 s + 3 s^2 - 8 t u)
\; ,
\quad  
\\[2mm]
\label{eq:c15}
  a_{C_5} &\! =\! &
 4 (8 m^4 - 3 s^2 + 2 t u)
\; ,
\quad  
\\[2mm]
\label{eq:c16}
\nonumber
  a_{C_6} &\! =\! &
- 4 \beta^2 (2 m^2 s + s^2 + 2 t u)
\; ,
\quad  
\\[2mm]
\label{eq:d11}
\nonumber
  a_{D_1} &\! =\! &
 4 ( 2 m^2 (2 D + s z_t \beta^2 - t^2 \beta^2)
+ s^2 t \beta^2 + t^3 )
\; ,   
\quad
\\[2mm]
\label{eq:d11u}
\nonumber
  a_{D_{1u}} &\! =\! &
   a_{D_1} (t\lra u)
\; ,   
\quad
\\[2mm]
\label{eq:d12}
\nonumber
  a_{D_2} &\! =\! &
 4 (8 m^2 D - s t u \beta^2 + 2 t^2/s (t^2 + u^2))
\; ,   
\quad
\\[2mm]
\label{eq:d12u}
\nonumber
  a_{D_{2u}} &\! =\! &
     a_{D_2} (t\lra u)
\; ,
\quad  
\\[2mm]
\label{eq:d13}
\nonumber
  a_{D_3} &\! =\! &
 8 (8 m^2 D - 8 m^4 t u/s - s t u \beta^2 + 2 t^2 u^2/s)
\; ;
\ea
\ba
\label{eq:b12ep}
  a^{(\ep)}_{B_2} &\! =\! &
4 (2 s - z_2^2/(s \beta^2) )
\nonumber
\; ,
\quad  
\\[2mm]
\label{eq:b15ep}
 a^{(\ep)}_{B_5} &\! =\! &
\nonumber
               -a^{(\ep)}_{B_2}
\; ,
\quad  
\\[2mm]
\label{eq:c11ep}
a^{(\ep)}_{C_1} &\! =\! &
\nonumber
2 ( \beta^2 (s^3 (8 m^2 + s) - 8 t^2 u^2)/D
\nonumber \\
& &\mbox{\hspn}
 + 16 m^2/s (s^2 + t u + D/\beta^2)\nonumber
\; ,
\quad
\\[2mm]  
\label{eq:c12ep}
\nonumber
  a^{(\ep)}_{C_2} &\! =\! &
- 4 t^2 (10 - t/s (2 t u \beta^2 + 2 s^2 - 3 t^2)/D)
\; ,
\quad
\\[2mm]  
\label{eq:c12uep}
a^{(\ep)}_{C_{2u}} &\! =\! &
\nonumber
   a^{(\ep)}_{C_2} (t\lra u)
\; ,
\quad
\\[2mm]
\label{eq:c13ep}
a^{(\ep)}_{C_{3}} &\! =\! &
\nonumber
 - 4 t^2 (6 - t/s (2 t u \beta^2 + 2 s^2 - 4 s t - 5 t^2)/D)
\; ,
\quad  
\\[2mm]
\label{eq:c13uep}
\nonumber
  a^{(\ep)}_{C_{3u}} &\! =\! &
  a^{(\ep)}_{C_{3}} (t\lra u)
\; ,
\quad  
\\[2mm]
\label{eq:c14ep}
\nonumber
  a^{(\ep)}_{C_4} &\! =\! &
- 2 ( s^3 (2 m^2 - s) - 8 t u (m^2 s + t^2 + u^2) )/D
\; ,
\quad  
\\[2mm]
\label{eq:c15ep}
  a^{(\ep)}_{C_5} &\! =\! &
 2 ( \beta^2 (s^3 (6 m^2 - s) + 4 t^2 u^2)/D
\\
& & \mbox{\hspn}
+ 8 s^2 - 12 m^4 z_2^2/D )\nonumber
\; ,
\quad
\\[2mm]
\label{eq:c16ep}
\nonumber
  a^{(\ep)}_{C_6} &\! =\! &
2 ( \beta^2 (s^3 (8 m^2 - s) - 4 t^2 u^2)/D
\nonumber \\
& & \mbox{\hspn} 
+ 8 m^2 s - 4 m^4 z_2^2/D )\nonumber
\; ,
\quad  
\\[2mm]
\label{eq:d11ep}
\nonumber
  a^{(\ep)}_{D_1} &\! =\! &
 - 2 t ( 2 s^2 \beta^2 - s^2 t \beta^2 (2 m^2 z_2^2/s^2 + 4 m^2  + t)/D 
\\
& &\mbox{\hspn}
                  + 2 t z_t + 2 s^2 )
\nonumber
\; , 
\quad  
\\[2mm]
\label{eq:d11uep}
\nonumber
  a^{(\ep)}_{D_{1u}} &\! =\! &
   a^{(\ep)}_{D_1} (t\lra u)        \nonumber
\; ,   
\quad  
\\[2mm]
\label{eq:d12ep}
\nonumber
  a^{(\ep)}_{D_2} &\! =\! &
-2 t ( 2 s (s - u) + t^2 (s^2 + 8 t u - 8 u^3/s)/D )
\; ,
\quad 
\\[2mm]
\label{eq:d12uep}
\nonumber
  a^{(\ep)}_{D_{2u}} &\! =\! &
   a^{(\ep)}_{D_2} (t\lra u)
\; ,
\quad
\\[2mm]
\label{eq:d13ep}
\nonumber
  a^{(\ep)}_{D_3} &\! =\! &
4 t u ( 4 s - t^2 u^2/s^2 (8 m^2 - 7 s)/D )
\; ;
\ea
\ba
\label{eq:b12eep}
  a^{(\ep^2)}_{B_2} &\! =\! &
0
\nonumber
\; ,
\quad  
\\[2mm]
\label{eq:b15eep}
 a^{(\ep^2)}_{B_5} &\! =\! &
\nonumber
               0
\; ,
\quad  
\\[2mm]
\label{eq:c11eep}
a^{(\ep^2)}_{C_1} &\! =\! &
\nonumber
 8 s (s - 2 m^2 z_2^2/D)
\; ,   
\quad
\\[2mm]
\label{eq:c12eep}
\nonumber
  a^{(\ep^2)}_{C_2} &\! =\! &
- 8 t^2 (3 u/s + t (m^2 - u)/D)
\; ,  
\quad
\\[2mm]
\label{eq:c12ueep}
\nonumber
  a^{(\ep^2)}_{C_{2u}} &\! =\! &  a^{(\ep^2)}_{C_2} (t\lra u)
\; ,
\quad  
\\[2mm]
\label{eq:c13eep}
a^{(\ep^2)}_{C_{3}} &\! =\! &
\nonumber
8 t^2 (2 + t u (1 - 3 t/s)/D)
\; ,
\quad  
\\[2mm]
\label{eq:c13ueep}
\nonumber
  a^{(\ep^2)}_{C_{3u}} &\! =\! &  a^{(\ep^2)}_{C_{3}} (t\lra u)
\; ,
\quad  
\\[2mm]
\label{eq:c14eep}
  a^{(\ep^2)}_{C_4} &\! =\! &
- 16 s^2 t u/D
\; ,
\quad  
\\[2mm]
\label{eq:c15eep}
\nonumber
  a^{(\ep^2)}_{C_5} &\! =\! &
- 4 s (s + 2 m^2 z_2^2/D)
\; ,
\quad  
\\[2mm]
\label{eq:c16eep}
\nonumber
  a^{(\ep^2)}_{C_6} &\! =\! &
4 s (s - 2 m^2 z_2^2/D)
\; ,
\quad  
\\[2mm]
\label{eq:d11eep}
\nonumber
  a^{(\ep^2)}_{D_1} &\! =\! &
- 4 s t ( 2 m^2 - s + t (\beta^2 t u + m^2 z_2^2/s)/D )
\; ,
\quad  
\\[2mm]
\label{eq:d11ueep}
\nonumber
  a^{(\ep^2)}_{D_{1u}} &\! =\! &  a^{(\ep^2)}_{D_1} (t\lra u)
\; ,
\quad  
\\[2mm]
\label{eq:d12eep}
\nonumber
  a^{(\ep^2)}_{D_2} &\! =\! &
 4 s t (s - 4 t^2 u/D)
\; ,
\quad  
\\[2mm]
\label{eq:d12ueep}
\nonumber
  a^{(\ep^2)}_{D_{2u}} &\! =\! &  a^{(\ep^2)}_{D_2} (t\lra u)
\; ,
\quad  
\\[2mm]
\label{eq:d13eep}
\nonumber
  a^{(\ep^2)}_{D_3} &\! =\! &
- 8 t u (s + 3 t^2 u^2/(s D))
\; ;
\end{eqnarray}

\begin{eqnarray}
\label{eq:con2}
  b_{1} &\! =\! &
- 16/3 z_2/s ( m^2 (n_l + 1) + (2 C_F - N_C) 3 D/(s \beta^2)
                                               \nn  \\   && \mbox{\hspn}
- N_C (m^2 + D 6 (10 m^2 - s)/(s^2 \beta^4 ) ) )
\nonumber
\; ,
\quad
\\[2mm]  
\label{eq:b22}
  b_{B_2} &\! =\! &
- 8 z_2/s^2 ( 8 m^4 - (2 C_F - N_C) D (2 + 1/\beta^2) )
\nonumber
\; ,
\quad
\\[2mm]  
\label{eq:b25}
 b_{B_5} &\! =\! &
\nonumber
               - N_C 8 z_2 ( D (16 m^2 - s)/(s \beta^4) + t u )/s^2
\; ,
\quad
\\[2mm]
\label{eq:c21}
b_{C_1} &\! =\! &
\nonumber
- N_C 16 m^2 D z_2 (8 m^2 + s)/(s^3 \beta^4)
\; ,   
\quad
\\[2mm]
\label{eq:c24}
  b_{C_4} &\! =\! &
 N_C 4 z_2 (D - 2 t u)/s
\; ,
\quad  
\\[2mm]
\label{eq:c25}
b_{C_{5}} &\! =\! &
\nonumber
 - 32 m^4 z_2/s
\; ,
\quad  
\\[2mm]
\label{eq:c26}
b_{C_{6}} &\! =\! &
\nonumber
- (2 C_F - N_C) 16 D z_2 (2 m^2 - s)/s^2
\; ;
\ea
\ba
\label{eq:con2ep}
  b^{(\ep)}_{1} &\! =\! &
16/3 \, z_2 ( t u (n_l + 1) + (2 C_F - N_C) 3 D/\beta^2       
\nn \\  && \mbox{\hspn}
                  - N_C (36 m^2 D/(s \beta^4) - t u (4 m^2 - 7 s)/(s \beta^2)) )/s^2
\nonumber
\; ,   
\quad
\\[2mm]
\label{eq:b22ep}
  b^{(\ep)}_{B_2} &\! =\! &
 8 z_2 ( 8 m^2 t u/s + (2 C_F - N_C) (2 t u - D/\beta^2) )/s^2
\nonumber
\; ,   
\quad
\\[2mm]
\label{eq:b25ep}
 b^{(\ep)}_{B_5} &\! =\! &
\nonumber
          N_C 8 z_2 ( 3 m^2 z_2^2/(s \beta^4) - 2 (D + 2 m^2 t u/s)/\beta^2 )/s^2
\;  ,
\quad
\\[2mm]
\label{eq:c21ep}
b^{(\ep)}_{C_1} &\! =\! &
\nonumber
N_C 16 m^2 z_2 (3 D/\beta^4 + 2 t u/\beta^2)/s^2
\; ,   
\quad
\\[2mm]
\label{eq:c24ep}
  b^{(\ep)}_{C_4} &\! =\! &
N_C 12 t u z_2/s
\; ,
\quad  
\\[2mm]
\label{eq:c25ep}
b^{(\ep)}_{C_{5}} &\! =\! &
\nonumber
32 m^2 t u z_2/s^2
\; ,
\quad  
\\[2mm]
\label{eq:c26ep}
b^{(\ep)}_{C_{6}} &\! =\! &
\nonumber
- (2 C_F - N_C) 16 t u z_2 (2 m^2 - s)/s^2
\; ;
\end{eqnarray}

\begin{eqnarray}
\label{eq:con3}
  c_{1} &\! =\! &
16 ( C_F ( D \beta^2 (8 m^2 T/t^2 + 2)
                    - D (6 z_t/t - 2 - t/s)
\nn \\ && \mbox{\hspn}
                    + 2 m^2 (4 z_t (m^2/s - 1) - m^2)
                    - D (1 + 4 t/s)/\beta^2 )
\nonumber \\
& & \mbox{\hspn}
             -   N_C ( D 2 m^2 (2 D + t u)/(st^2)
                    - 2 m^2 t u/s
\nn \\ && \mbox{\hspn}
                    - D 4 m^2 (s + 4 t)/(s^2 \beta^2) ) )  /T
\nonumber
\; ,   
\quad
\\[2mm]
\label{eq:b31}
  c_{B_1} &\! =\! &
 16 ( C_F ( 2 m^2 \beta^2 (T - 2 s - D (2 T + t)/t^2) 
\nn \\ && \mbox{\hspn}
                       + D (3 z_t/t + t/s) - 2 m^2 u (2 + 5 t/s) )/T
\nn \\ && \mbox{\hspn}
         + N_C 2 D (D/s - t)/t^2 )
\nonumber
\; ,
\quad  
\\[2mm]
\label{eq:b32}
 c_{B_2} &\! =\! &
(2 C_F - N_C) 16 D/(s \beta^2)
\; ,
\quad
\\[2mm]
\label{eq:b35}
c_{B_5} &\! =\! &
\nonumber
N_C 8 ( - 8 m^2 D/(s^2 \beta^2) - t \beta^2 + t^2 z_2/s^2 )
\; ,
\quad
\\[2mm]
\label{eq:c31}
\nonumber
  c_{C_1} &\! =\! &
 N_C 8 ( t^3 + u^3 - 4 t^2 T - s D/\beta^2 - s^2 \beta^2 (m^2 - t) )/s
\; ,
\quad  
\\[2mm]
\label{eq:c32}
c_{C_{2}} &\! =\! &
\nonumber
- (2 C_F - N_C) 16 ( 2 m^2 z_2 (m^2 s/t - z_t) 
\\ \nn && \mbox{\hspn}
              + t (s^2 + t^2) )/s
\; ,
\quad  
\\[2mm]
\label{eq:c33}
c_{C_{3}} &\! =\! &
\nonumber
N_C 16 t ( 4 D/s - t \beta^2 + s )
\; ,
\quad  
\\[2mm]
\label{eq:c34}
c_{C_{4}} &\! =\! &
\nonumber
 N_C 4 ( - s^2 \beta^2 + 3 z_2 (m^2 s - t^2)/s - 3 s u + 2 t^2 )
\; ,
\quad  
\\[2mm]
\label{eq:c35}
c_{C_{5}} &\! =\! &
\nonumber
(2 C_F - N_C) 8 (2 T (2 m^2 + s) - u^2)
\; ,
\quad  
\\[2mm]
\label{eq:c36}
c_{C_{6}} &\! =\! &
\nonumber
- (2 C_F - N_C) 8 (4 m^2 D/s - 4 m^2 t \beta^2 + 3 t z_t - z_2^2)
\; ,
\quad  
\\[2mm]
\label{eq:d31}
c_{D_{1}} &\! =\! &
\nonumber
- (2 C_F - N_C) 8 ( m^2 s^2 \beta^4 - 2 m^2 t \beta^2 (s - t)
                                                      + s t^2 \beta^2 
\nn \\ && \mbox{\hspn}
                               - t^3 - s D )
\nn
\; ,   
\quad
\\[2mm]
\label{eq:d32}
c_{D_{2}} &\! =\! &
\nonumber
 N_C 8 (8 m^2 D - s t u \beta^2 + 2 t^2 (t^2 + u^2)/s)
\; ;
\ea
\ba
\label{eq:con3ep}
  c^{(\ep)}_{1} &\! =\! &
16 ( C_F ( D (16 m^2 D/(s t^2) - 24 m^4/t^2 + 4 - t/s 
\nn \\ && \mbox{\hspn}
                                                      + 2 t/(s \beta^2))
                        + 2 m^2 (4 m^2 - 6 t - 9 t^2/s)
\nn \\ && \mbox{\hspn}
                        + 4 m^2 t^2 z_2/(s^2 \beta^2) )/T
\nonumber \\
& &  \mbox{\hspn}
\nonumber
                 + N_C 2 ( 2 m^4 s/t^2 + t + D z_2/(s t)
\nn \\ && \mbox{\hspn}
                        - D (4 m^2 + 3 s)/(s^2 \beta^2) + t z_2/(s \beta^2) ) )
\nonumber
\; , 
\quad  
\\[2mm]
\label{eq:b31ep}   
  c^{(\ep)}_{B_1} &\! =\! &
16 (C_F (4 m^4 D/t^2 - 6 T D/t - 2 m^2 D/s - t D/s 
\nn \\ && \mbox{\hspn}
                                                   - 5 m^2 z_t + t^2)
\nn \\ && \mbox{\hspn}
                    - N_C (2 m^2 D^2/(s t^2) + 2 t D/s - m^4 + t^2))/T
\nonumber
\; ,   
\quad
\\[2mm]
\label{eq:b32ep}
 c^{(\ep)}_{B_2} &\! =\! &
 - (2 C_F - N_C) 8 t (2 + z_2/(s \beta^2))
\; ,
\quad
\\[2mm]  
\label{eq:b35ep}
c^{(\ep)}_{B_5} &\! =\! &
\nonumber
N_C 8 (2 D/\beta^2 + 6 m^2 z_2/\beta^2 - 3 t^2 - 2 t^3/s)/s
\; ,
\quad
\\[2mm]
\label{eq:c31ep}
\nonumber
  c^{(\ep)}_{C_1} &\! =\! &
- N_C 4 ( 2 m^2 z_2^2/s - 4 s^2 - 4 t^2 - 4 m^2 (4 z_t 
\nn \\ && \mbox{\hspn}
                                                       + t z_2/s)/\beta^2
               + 2 t u z_t (4 s + 3 t^2/s + u^2/s)/D 
\nn \\ && \mbox{\hspn}
+ s t^2 (2 t \beta^2 - z_2)/D )
\nn
\; ,   
\quad
\\[2mm]
\label{eq:c32ep}
c^{(\ep)}_{C_{2}} &\! =\! &
\nonumber
(2 C_F - N_C) 8 ( \beta^2 t (6 s D - 4 m^2 t u - s t^2)
\nn \\ && \mbox{\hspn}
                              + 2 D (2 m^4 s/t - s z_t + D t/s - t^2)
\nn \\ && \mbox{\hspn}
                              - 4 m^2 t^3 z_2/s )/D
\nn
\; ,
\quad
\\[2mm]
\label{eq:c33ep}
c^{(\ep)}_{C_{3}} &\! =\! &
\nonumber
- N_C 8 t^2 (4 s/t + 14 - s t (4 \beta^2 - 8 t T/s^2 + 5)/D)
,
\quad
\\[2mm]
\label{eq:c34ep}
c^{(\ep)}_{C_{4}} &\! =\! &
\nonumber
- N_C 4 (2 s^2 + 2 t^3/s - 2 s u + m^2 s t (9 s + 7 t)/D 
\\ &&   \nn \mbox{\hspn}
                                                         - t^4 (9 + 8 t/s)/D)
\; ,
\quad
\\[2mm]
\label{eq:c35ep}
c^{(\ep)}_{C_{5}} &\! =\! &
\nonumber
- (2 C_F - N_C) 4 (2 m^2 z_2^2/s - 2 t^2 
\\ && \nn \mbox{\hspn}
                                - 4 s \beta^2 (s - m^2 t u/D)
                                                          + t^2 (8 m^2 t + s^2)/D)
\; ,   
\quad
\\[2mm]
\label{eq:c36ep}
c^{(\ep)}_{C_{6}} &\! =\! &
\nonumber
- (2 C_F - N_C) 4 s t (7 \beta^2 + 2 u^2 \beta^2/(s t) + 2 s/t + 5
\\ &&  \nn \mbox{\hspn}
                                         + \beta^2 (2 m^2 z_2^2/s - 3 s t - 4 t^2)/D)
\; ,
\quad
\\[2mm]
\label{eq:d31ep}
c^{(\ep)}_{D_{1}} &\! =\! &
\nonumber
- (2 C_F - N_C) 4 s t^2 (2 s \beta^2/t + 2 s/t + 2 t/s 
\\ &&  \nn \mbox{\hspn}
                                                  + 4 m^4 z_2^2/(s^2 D)
\\ &&  \nn \mbox{\hspn}
                                          - \beta^2 (6 m^2 s - 4 m^2 t u/s + s t)/D)
 ,
\quad
\\[2mm]
\label{eq:d32ep}
c^{(\ep)}_{D_{2}} &\! =\! &
\nonumber
 - N_C 4 t (4 s^2 + 2 s t + t^2 (z_2^2 + 12 t u - 8 u^3/s)/D)
\; ;
\ea
\ba
\label{eq:con3eep}
  c^{(\ep^2)}_{1} &\! =\! &
- 16 ( C_F ( 2 m^2 (6 D/t + 4 m^2 + t)
\nn \\ &&
                           - 4 m^2 (3 D + t z_2)/(s \beta^2) + D/\beta^2 )/T
\nn \\ &&
                      - N_C 2 (u - 4 m^2 z_u/(s \beta^2)) )
\nonumber
\; ,
\quad
\\[2mm]
\label{eq:b31eep}
  c^{(\ep^2)}_{B_1} &\! =\! &
16 ( C_F z_t (3 D/t + 2 m^2)/T - N_C 2 m^2 )
\nonumber
\; ,
\quad
\\[2mm]
\label{eq:b32eep}
 c^{(\ep^2)}_{B_2} &\! =\! &
\nonumber
- (2 C_F - N_C) 32 m^2 z_2/(s \beta^2)
\; ,
\quad
\\[2mm]
\label{eq:b35eep}
c^{(\ep^2)}_{B_5} &\! =\! &
\nonumber
- N_C 64 m^2 z_2/(s \beta^2)
\; ,   
\quad
\\[2mm]
\label{eq:c31eep}
\nonumber
  c^{(\ep^2)}_{C_1} &\! =\! &
- N_C 8 s (4 t + 2 m^2 t (s + 4 t + z_2^2/(s \beta^2))/D 
\\ &&  \nn \mbox{\hspn}
                                                         + s/\beta^2 )
\; ,
\quad  
\\[2mm]
\label{eq:c32eep}
c^{(\ep^2)}_{C_{2}} &\! =\! &
\nonumber
(2 C_F - N_C) 16 ( 2 m^2 s + t u + 2 m^2 t (m^2 s 
\\ &&  \nn \mbox{\hspn}
                                                  - t^2)/D )
\; ,
\quad  
\\[2mm]
\label{eq:c33eep}
c^{(\ep^2)}_{C_{3}} &\! =\! &
\nonumber
N_C 16 t (s + 2 t (m^2 s + t u)/D)
\; ,
\quad  
\\[2mm]
\label{eq:c34eep}
c^{(\ep^2)}_{C_{4}} &\! =\! &
N_C 8 s (s + 2 t (m^2 s + t u)/D)
\; ,
\quad  
\\[2mm]
\label{eq:c35eep}
c^{(\ep^2)}_{C_{5}} &\! =\! &
\nonumber
(2 C_F - N_C) 8 s (u - 2 m^2 t z_2/D )
\; ,
\quad  
\\[2mm]
\label{eq:c36eep}
c^{(\ep^2)}_{C_{6}} &\! =\! &
\nonumber
- (2 C_F - N_C) 8 ( 2 m^2 z_2 (m^2 s - t^2)/D + s u )
\; ,
\quad  
\\[2mm]
\label{eq:d31eep}
c^{(\ep^2)}_{D_{1}} &\! =\! &
\nonumber
- (2 C_F - N_C) 8 t (s z_u + 2 m^2 t z_2^2/D)
\; ,
\quad  
\\[2mm]
\label{eq:d32eep}
c^{(\ep^2)}_{D_{2}} &\! =\! &
\nonumber
N_C 8 s t (s - 4 t^2 u/D)
\; .
\end{eqnarray}

\section{}

In this Appendix, we present the expressions for all the coefficients 
$d_I, e_I, g_I, h_I$
appearing in Eq.~(\ref{F6}):

\ba
\label{eq:db2}
  d_{B_2}^{(\ep)} &\! =\! &
  2 s ( 4 m^2 + z_2^2/(s \beta^2) )/(t u)
\nonumber
\; ,   
\quad
\\[2mm]
\label{eq:db5} 
  d_{B_5}^{(\ep)} &\! =\! &
  - d_{B_2}^{(\ep)}
\nonumber
\; ,  
\quad
\\[2mm]  
\label{eq:dc1} \nn
  d_{C_1}^{(\ep)} &\! =\! &
  s ( s^2 \beta^2 (8 m^2 s + s^2 + 2 D) + 4 s^2 D
\\ &&  \nn \mbox{\hspn}
                 - 16 m^2 D^2/(s\beta^2) - 8 m^4 z_2^2 )/(t u D)
\nonumber 
\; ,
\quad
\\[2mm]
\label{eq:dc2}
  d_{C_2}^{(\ep)} &\! =\! &
  2 t ( 2 u (D + m^2 s) + s t^2 + {\kappa}_{d} t/s )/(u D)
\nonumber
\; ,
\quad 
\\[2mm]
\label{eq:dc2u}
  d_{C_{2u}}^{(\ep)} &\! =\! &
    d_{C_2}^{(\ep)} (t\lra u)
\nonumber
\; ,
\quad  
\\[2mm]
\label{eq:dc3}
  d_{C_3}^{(\ep)} &\! =\! &
    - 2 t ( 2 m^2 s^2 + s t^2 - \kappa_d t/s )/(u D)
\nonumber
\; ,
\quad
\\[2mm]
\label{eq:dc13u}
  d_{C_{3u}}^{(\ep)} &\! =\! &
    d_{C_3}^{(\ep)} (t\lra u)
\; , 
\quad  
\\[2mm]
\label{eq:dc14} 
  d_{C_4}^{(\ep)} &\! =\! &
    s^2 ( \kappa_d + 3 s D - s^2 (m^2 - s) )/(t u D)
\nonumber
\; ,
\quad
\\[2mm]
\label{eq:dc15} 
  d_{C_5}^{(\ep)} &\! =\! &
    - s ( \kappa_c + 2 m^2 s z_2^2 )/(t u D)
\nonumber
\; ,
\quad
\\[2mm]
\label{eq:dc16}
  d_{C_6}^{(\ep)} &\! =\! &
    - s \kappa_c/(t u D)  
\nonumber
\; ,
\quad
\\[2mm]
\label{eq:dd11}
  d_{D_1}^{(\ep)} &\! =\! &
    s t ( z_2 + \beta^2 (\kappa_d - s^2 (m^2 - t))/D )/u  
\nonumber
\; , 
\quad  
\\[2mm]
\label{eq:dd11u}
  d_{D_{1u}}^{(\ep)} &\! =\! &
    d_{D_1}^{(\ep)} (t\lra u)
\nonumber
\; ,
\quad
\\[2mm]
\label{eq:dd12}
  d_{D_2}^{(\ep)} &\! =\! &
    s t ( \kappa_d + s D + s^2 (m^2 - t) )/(u D)
\nonumber
\; ,
\quad
\\[2mm]
\label{eq:dd12u} 
  d_{D_{2u}}^{(\ep)} &\! =\! &
    d_{D_2}^{(\ep)} (t\lra u)  
\nonumber
\; ,
\quad
\\[2mm]
\label{eq:dd13}
  d_{D_3}^{(\ep)} &\! =\! &   
    2 t u \kappa_d/(s D)
\nonumber
\; ,
\quad
\\[2mm]
{\rm with} \nn \\ \nn
\kappa_c &\! =\! & 4 D^2 - s \beta^2 (8 m^2 s^2 - 8 m^2 t u - s^3) \; ,    \\ \nn
\kappa_d &\! =\! & 10 m^2 s^2 - 8 m^2 t u - 3 s t u \; ;
\ea

\ba
\label{eq:db2e}
  d_{B_2}^{(\ep^2)} &\! =\! &
    - 8 m^2 z_2^2/(s t u \beta^2)
\nonumber
\; ,   
\quad  
\\[2mm]
\label{eq:db5e}
  d_{B_5}^{(\ep^2)} &\! =\! &
  - d_{B_2}^{(\ep^2)}
\nonumber
\; ,   
\quad
\\[2mm]
\label{eq:dc1e}
  d_{C_1}^{(\ep^2)} &\! =\! &
  - s ( (22 m^2 s^2 - 16 m^2 t u + s^3) s \beta^2
                                 + 4 m^2 s D      \nn
\\ &&  \nn \mbox{\hspn}
- 16 m^2 D^2/(s \beta^2) )/(t u D)
\; ,   
\quad
\\[2mm]
\label{eq:dc2e}
  d_{C_2}^{(\ep^2)} &\! =\! &
  2 t ( 6 s D + 4 m^2 s z_2 + t^2 z_2 - \kappa_d t/s )/(u D)
\nonumber
\; ,
\quad  
\\[2mm]                            \nn \\
\label{eq:dc2ue}
  d_{C_{2u}}^{(\ep^2)} &\! =\! &
    d_{C_2}^{(\ep^2)} (t\lra u)
\nonumber
\; ,
\quad  
\\[2mm]
\label{eq:dc3e}
  d_{C_3}^{(\ep^2)} &\! =\! &
    - 2 t ( \kappa_d t/s - s ( 2 m^2 s - 2 s t - t^2) )/(u D)
\nonumber
\; ,   
\quad  
\\[2mm]
\label{eq:dc13ue}
  d_{C_{3u}}^{(\ep^2)} &\! =\! &
    d_{C_3}^{(\ep^2)} (t\lra u)
\; ,   
\quad
\\[2mm]
\label{eq:dc14e}
  d_{C_4}^{(\ep^2)} &\! =\! &
    - s^2 ( \kappa_d + s D + s^2 (m^2 + s) )/(t u D)
\nonumber
\; ,   
\quad
\\[2mm]
\label{eq:dc15e}
  d_{C_5}^{(\ep^2)} &\! =\! &
    - s ( \kappa_c - 2 m^2 s ( 4 D + z_2^2) )/(t u D)
\nonumber
\; ,   
\quad
\\[2mm]
\label{eq:dc16e}
  d_{C_6}^{(\ep^2)} &\! =\! &
    - s \kappa_c/(t u D)
\nonumber
\; ,   
\quad
\\[2mm]
\label{eq:dd11e}
  d_{D_1}^{(\ep^2)} &\! =\! &
    s t ( 2 u D - \beta^2 (\kappa_d - 4 s t u - s t^2) )/(u D)
\nonumber
\; ,   
\quad
\\[2mm]
\label{eq:dd11ue}
  d_{D_{1u}}^{(\ep^2)} &\! =\! &
    d_{D_1}^{(\ep^2)} (t\lra u)
\nonumber
\; ,   
\quad
\\[2mm]
\label{eq:dd12e}
  d_{D_2}^{(\ep^2)} &\! =\! &
    - s t ( \kappa_d + s (2 m^2 s - 2 s t - t^2) )/(u D)
\nonumber
\; ,   
\quad
\\[2mm]
\label{eq:dd12ue}
  d_{D_{2u}}^{(\ep^2)} &\! =\! &
    d_{D_2}^{(\ep^2)} (t\lra u)
\nonumber
\; ,   
\quad
\\[2mm]
\label{eq:dd13e}
  d_{D_3}^{(\ep^2)} &\! =\! &
    - 2 t u \kappa_d/(s D)
\nonumber
\; ,   
\quad
\\[2mm]
{\rm with} \nn \\ \nn
\kappa_c &\! =\! & 4 t u D + s\beta^2 (18 m^2 s^2 - 16 m^2 t u - s^3) \; ,    \\ \nn
\kappa_d &\! =\! & 18 m^2 s^2 - 16 m^2 t u + s t u  \; ;
\ea

\ba
  e_{1}^{(\ep)} &\!=\!&   2 s (n_l + 1) \kappa_1 \kappa_2     \; ,  
\qquad
  e_{B_2}^{(\ep)} =  3 (8 m^2 + s) \kappa_1 \kappa_2
\nonumber
\; ,
\quad
\\[2mm]  
  e_{B_5}^{(\ep)} &\!=\!&   3 s n_l \kappa_1 \kappa_2     \; , 
\qquad
  e_{C_5}^{(\ep)} =   18 m^2 s \kappa_1 \kappa_2
\nonumber
\; ,
\quad
\\[2mm]
%
  e_{1}^{(\ep^2)} &\! =\! &    e_{1}^{(\ep)}/ \kappa_2   \; ,
\qquad
  e_{B_2}^{(\ep^2)} =  e_{B_2}^{(\ep)}/ \kappa_2
\; ,
\quad
\\[2mm]
  e_{B_5}^{(\ep^2)} &\! =\! &    e_{B_5}^{(\ep)}/ \kappa_2  \; ,   
\qquad
  e_{C_5}^{(\ep^2)} =   e_{C_5}^{(\ep)}/ \kappa_2
\nonumber
\; ,
\quad
\\[2mm]
{\rm with} \nn \\ \nn
\kappa_1 &\! =\! & 8 z_2/(9 s^2)   \; ,    \\ \nn
\kappa_2 &\! =\! & - m^2 s/(t u)    \; .
\ea
%
Next, we introduce common factors that appear in the various coefficients $g_I$
and $h_I$. They are multiplied by one power of $\ep$ and read
\ba
\label{scommons}
s_{b2} &\!=\!&  2 m^2 s - t z_2^2/(s\beta^2) \; ,        \\
s_{c2} &\!=\!&       t s_{c5} - 4 m^2 s u D   \;  ,      \nn  \\
s_{c5} &\!=\!& 2 D (D + s (8 m^2 + t)) + 2 s t \beta^2 (2 m^2 u - t^2) + s t^2 z_2  \; ,
\nn \\  \nn
s_{c6} &\!=\!& D (3 s \beta^2 + z_2) + \beta^2 (6 m^2 s^2 - 8 m^2 t u + s^2 t)   \;  .
\ea
For the coefficients $g_I^{(\ep)}$, we have
\ba
\label{eq:gcon}
  g_{1}^{(\ep)} &\! =\! &
    8 ( 2 m^2 s \beta^2 (4 s T^2/t + t z_t - 2 t u)
              + D (10 m^2 u - 5 s z_t       \nn   \\   && \mbox{\hspn}
- 2 t^2) - 2 m^2 t (s^2 + u^2)
                 - D^2 6 t/(s \beta^2)      \nn   \\   && \mbox{\hspn}
- 3 D t^2 z_2/(s\beta^2) )/(t^2 u T)
\nonumber
\; ,
\quad  
\\[2mm]
\label{eq:gb1}
  g_{B_1}^{(\ep)} &\! =\! &
      - 8 ( D (4 m^2 u - s t) - t^3 (2 s\beta^2 + 3 z_t) )/(t^2 u T)
\nonumber
\; ,
\quad  
\\[2mm]
g_{B_2}^{(\ep)} &\! =\! &   - 8 s_{b2}/(t u)  \;  ,
\qquad
g_{C_2}^{(\ep)}  =  8 s_{c2}/(D t u)  \;  , 
\quad
\\[2mm]
g_{C_5}^{(\ep)} &\! =\! &   4 s s_{c5}/(D t u)    ,
\quad
g_{C_6}^{(\ep)} = 4 s s_{c6}/(D u)     ,
\quad
g_{D_1}^{(\ep)} =  - t g_{C_6}^{(\ep)}.
\nn
\ea
Finally, we introduce factors that are common to various coefficients $g_I$
and $h_I$ that are multiplied by two powers of $\ep$:
\ba
\label{ccommons}
c_{b2} &\!=\!&  2 m^2 u + D  \; ,        \\
c_{c2} &\!=\!&  2 D ( 4 m^2 s u - t D - s t (17 m^2 + 3 t) )
                                                      \nn   \\ && \mbox{\hspn}
    - 4 s t^2 \beta^2 (2 m^2 u - t^2) + s t^2 (3 s z_t + 4 m^2 z_2)   \;  ,  
\nn  \\
c_{c5} &\!=\!& 2 D (20 m^2 s - s t + t^2) + 2 s t \beta^2 ( 4 m^2 u + s t - 2 t^2) 
                                                       \nn   \\ && \mbox{\hspn}
                       + 5 s t^2 z_2  \; ,
\nn \\  \nn
c_{c6} &\!=\!& 2 D (s\beta^2 - u) + \beta^2 (16 m^2 (s^2 - t u) + s^2 t)   \;  .
\ea
For the coefficients $g_I^{(\ep^2)}$, we have
\ba
\label{eq:gcone}
  g_{1}^{(\ep^2)} &\! =\! &
8 ( D^2 12 t/(s\beta^2) + D^2 16 m^2/t
           + D 4 (2 s z_t - t u)               \nn  \\  && \mbox{\hspn}
- D t (14 m^2 + 3 t)/\beta^2
           - 2 m^2 (12 m^2 s^2 T/t + 5 t^3)     \nn  \\  && \mbox{\hspn}
- 4 m^2 t^2 z_2/\beta^2 )/(t^2 u T)
\nonumber
\; ,
\quad
\\[2mm]
\label{eq:gb1e}
  g_{B_1}^{(\ep^2)} &\! =\! &
      8 ( D (4 z_t/t^2 + 1/u) + 2 m^2 (t/u - 2) )/T
\nonumber
\; ,
\quad
\\[2mm]
g_{B_2}^{(\ep^2)} &\! =\! &   - 16 z_2 c_{b2}/(s t u\beta^2)   \;  ,
\qquad
g_{C_2}^{(\ep^2)}  =  8 c_{c2}/(D t u)   \;  ,   \nn
\quad
\\[2mm]
g_{C_5}^{(\ep^2)} &\! =\! &   - 4 s c_{c5}/(D t u)  \;  ,
\qquad 
g_{C_6}^{(\ep^2)} = - 4 s c_{c6}/(D u)  \; ,
\nn
\quad
\\[2mm]
g_{D_1}^{(\ep^2)} &\! =\! &  - t g_{C_6}^{(\ep^2)}  \; .
\ea
For the remaining coefficients $h_I$, we get
\ba  
\label{eq:hcon}
  h_{1}^{(\ep)} &\! =\! &
        8 ( t z_2 (2 m^2 + D/(s\beta^2))/\beta^2 - m^2 (4 s D/t - 2 s^2 
                           \nn \\ && \mbox{\hspn}
                                   + t z_2/9) )/(t^2 u)
\nonumber
\; ,
\quad
\\[2mm]
\label{eq:hb1}
  h_{B_1}^{(\ep)} &\! =\! &
      - 8 (3 m^2 s + t z_2)/(t u)
\nonumber
\; ,
\quad
\\[2mm]
  h_{B_2}^{(\ep)} &\! =\! &   4 s_{b2}/(t u)  \; ,
\quad
\\[2mm]
   h_{B_5}^{(\ep)} &\! =\! & 
    4 ( 2 D (8 m^2 t + s^2)/(s^2\beta^4) + 4/3 m^2 (s - u) 
                            \nn  \\  && \mbox{\hspn}
                                           - s z_t/\beta^2 )/(t u) \; ,
\nn 
\quad \\[2mm]
   h_{C_1}^{(\ep)} &\! =\! &
       - 2 s ( s^2\beta^4 t/D + 2 s\beta^2 (2 m^2 z_2 - t u)/D 
                            \nn  \\  && \mbox{\hspn}
- 8 m^2 s^2 z_t/(t D)
            - D 6/t + 8 t + 2 t z_2/(s\beta^2) 
                             \nn  \\  && \mbox{\hspn}
             - D 4 z_t/(s t\beta^4) )/u       \; ,
\nn
\quad \\[2mm]
    h_{C_2}^{(\ep)} &\! =\! &   - 4 s_{c2}/(D t u) \; ,
\qquad \nn
    h_{C_3}^{(\ep)} = 2 t/s h_{C_4}^{(\ep)}  \;  ,
\quad \\[2mm]
    h_{C_4}^{(\ep)} &\! =\! & 
2 s ( 8 m^2 u^2/D + 4 s - s t (8 m^2 + s)/D )/u   \;  ,
\nn
\quad \\[2mm]
   h_{C_5}^{(\ep)} &\! =\! &   - 2 s s_{c5}/(D t u) \; ,
\qquad \nn
   h_{C_6}^{(\ep)} = - 2 s s_{c6}/(D u)  \;  ,
\quad \\[2mm]
   h_{D_1}^{(\ep)} &\! =\! & - t h_{C_6}^{(\ep)}  \;  ,
\qquad \nn
   h_{D_2}^{(\ep)} = - t h_{C_4}^{(\ep)} \; ; 
\ea
\ba
\label{eq:hcone}
  h_{1}^{(\ep^2)} &\! =\! &
        8 ( 4 m^4 s^2/t - s^2 z_t - 10/9 s t^2 - t^3/3 - 2/9 t^4/s
                \nn \\ && \mbox{\hspn}
                   - D t z_2/(s\beta^4) + t (u z_2 + 8 m^2/s D)/\beta^2 )/(t^2 u)
\nonumber
\; ,
\quad
\\[2mm]
\label{eq:hb1e}
  h_{B_1}^{(\ep^2)} &\! =\! &
      16 (D z_2 + t^2 u)/(t^2 u)
\nonumber 
\; ,
\quad
\\[2mm]
  h_{B_2}^{(\ep^2)} &\! =\! &   8 z_2 c_{b2}/(s t u\beta^2)  \; ,
\quad
\\[2mm]
   h_{B_5}^{(\ep^2)} &\! =\! &
    - 8 z_2 ( 4 m^2 D/(s^2\beta^4) + t u/(3 s) 
                    \nn \\ && \mbox{\hspn}
                                         - 2 m^2 u/(s\beta^2) )/(t u) \; ,
\nn
\quad \\[2mm]
   h_{C_1}^{(\ep^2)} &\! =\! &
- 2 s ( 40 m^2 s/t
                 + (8 m^2 s u\beta^2 + 20 m^2 s u + 16 m^2 t^2 
   \nn \\ && \mbox{\hspn}
                                                                - s^2 t)/D
                 + 2 (2 m^2 s^2/t + 4 m^2 t - s^2)/(s\beta^2)
   \nn \\ && \mbox{\hspn}
                 + 8 m^2 D/(s t \beta^4) + 4 m^2 (1/s^2 + \beta^2/D) z_2^2/\beta^4 )/u ,
\nn
\quad \\[2mm]  
    h_{C_2}^{(\ep^2)} &\! =\! &   - 4 c_{c2}/(D t u) \; ,
\qquad \nn
    h_{C_3}^{(\ep^2)} = 2 t/s h_{C_4}^{(\ep^2)}  \;  ,
\quad \\[2mm]
    h_{C_4}^{(\ep^2)} &\! =\! &
- 2 s ( 20 m^2 s^2 + 4 s t u\beta^2 + s t z_2 )/(D u)   \;  ,
\nn
\quad \\[2mm]
   h_{C_5}^{(\ep^2)} &\! =\! &   2 s c_{c5}/(D t u) \; ,
\qquad \nn
   h_{C_6}^{(\ep^2)} =  2 s c_{c6}/(D u)  \;  ,
\quad \\[2mm]
   h_{D_1}^{(\ep^2)} &\! =\! & - t h_{C_6}^{(\ep^2)}  \;  ,
\qquad \nn
   h_{D_2}^{(\ep^2)} = - t h_{C_4}^{(\ep^2)} \; .
\ea

\section{}

In this Appendix, we present the expressions for all the coefficients 
$k_I, l_I, m_I, n_I$
appearing in Eq.~(\ref{F10}) using the following abbreviations:
\ba  \nn
\kappa_{c1} &\!=\!& 
-s \beta^2 (18 m^2 s^2 - s^3 + 2 (2 m^2 + s) z_2^2) - 8 m^2 s D   \; ,  \\
\kappa_{c6} &\!=\!&  -s \beta^2 z_2^2 + 16 s D + 6 s t u  \; .
\ea
We have
\ba
\label{eq:kc1}
  k_{C_1}^{(\ep^2)} &\! =\! &
        - s \kappa_{c1}/(t u D)
\nonumber
\; ,
\quad  
\\[2mm]
\label{eq:kc2}
  k_{C_2}^{(\ep^2)} &\! =\! &  
        2 t ( \kappa_{c6} t/s + 2 m^2 s u - s^2 z_t + t^3 )/(u D)
\nonumber
\; ,
\quad
\\[2mm]   
  k_{C_{2u}}^{(\ep^2)} &\! =\! &    k_{C_2}^{(\ep^2)} (t\lra u)  
\nonumber
\; ,
\quad
\\[2mm]
\label{eq:kc3}
  k_{C_3}^{(\ep^2)} &\! =\! &  
        2 t^2 ( \kappa_{c6}/s + 2 s^2 + u^2 )/(u D)
\nonumber
\; ,
\quad
\\[2mm]   
  k_{C_{3u}}^{(\ep^2)} &\! =\! &    k_{C_3}^{(\ep^2)} (t\lra u)
\; ,
\quad
\\[2mm]
\label{eq:kc4}
  k_{C_4}^{(\ep^2)} &\! =\! &
        s^2 ( \kappa_{c6} + 2 s (u^2 - s t) )/(t u D)
\nonumber
\; ,
\quad 
\\[2mm]
\label{eq:kc5}
  k_{C_5}^{(\ep^2)} &\! =\! &
        - s ( \kappa_{c1} - 2 s^2 (2 D - t u\beta^2 - z_2^2) )/(t u D) 
\nonumber
\; ,
\quad
\\[2mm]
\label{eq:kc6}
  k_{C_6}^{(\ep^2)} &\! =\! &
        s^2\beta^2 \kappa_{c6}/(t u D)  
\nonumber
\; ,
\quad
\\[2mm]
\label{eq:kd1}
  k_{D_1}^{(\ep^2)} &\! =\! &
        s t \beta^2 ( \kappa_{c6} - s u z_2 )/(u D)  
\nonumber
\; ,
\quad
\\[2mm]
  k_{D_{1u}}^{(\ep^2)} &\! =\! &   k_{D_1}^{(\ep^2)} (t\lra u)  \; ,
\nn \quad \\ \nn \\
\label{eq:kd2} 
  k_{D_2}^{(\ep^2)} &\! =\! &
        s t ( \kappa_{c6} + s^3 - s^2 t )/(u D)  
\nonumber
\; ,
\quad
\\[2mm]
  k_{D_{2u}}^{(\ep^2)} &\! =\! &  k_{D_2}^{(\ep^2)} (t\lra u) \; ,
\nn \quad \\[2mm]
\label{eq:kd3}
  k_{D_3}^{(\ep^2)} &\! =\! &
        2 t u ( \kappa_{c6}/s - s t + u^2 )/D  
\nonumber
\; ;
\ea
\ba
  l_1^{(\ep^2)} &\!=\!&    4/3 s (n_l + 1) \kappa_1\kappa_2  \;  ,
\qquad
  l_{B_2}^{(\ep^2)} =  (16 m^2 + 5 s) \kappa_1\kappa_2  \;  ,
\nn
\\[2mm]
  l_{B_5}^{(\ep^2)} &\!=\!&  5 s n_l \kappa_1\kappa_2  \;  ,
\qquad
  l_{C_5}^{(\ep^2)} =  18 m^2 s \kappa_1\kappa_2  \;  ;
\ea
\ba
\label{eq:m1}
  m_{1}^{(\ep^2)} &\! =\! &
        32 ( 2 m^2 (2 m^2 u^2/t^2 - s^2/t + 2 t - 8 D T/t^2) 
\nn \\ &&
                  - s\beta^2 (2 D/t + m^2) + D 4 U/(s\beta^2) )/(t u)
\nonumber
\; ,   
\quad
\\[2mm]
  m_{C_2}^{(\ep^2)} &\! =\! &
        2 z_t/(s\beta^2) m_{C_6}^{(\ep^2)}
\nonumber
\; ,
\quad
\\[2mm]
  m_{C_5}^{(\ep^2)} &\! =\! &
        z_t/(t\beta^2) m_{C_6}^{(\ep^2)}
\; ,  
\quad  
\\[2mm]
\label{eq:mc6}
  m_{C_6}^{(\ep^2)} &\! =\! &
        4 s\beta^2 ( \kappa_{c6}/u - s z_2 )/D
\nonumber
\; ,
\quad
\\[2mm]
\nn
   m_{D_1}^{(\ep^2)} &\! =\! &   - t m_{C_6}^{(\ep^2)}  \;  ;
\ea
\ba
\label{eq:n1}
  n_{1}^{(\ep^2)} &\! =\! &
         - 16 m^2 ( 18 s^2 z_t/t^2 + 82/3 s + 2/3 t
                  - 9 s z_2/(s\beta^2) 
\nn  \\  &&
                        + 144 z_u D/(s^2\beta^4) )/(9 t u)
\nonumber
\; , 
\quad  
\\[2mm]
  n_{B_5}^{(\ep^2)} &\! =\! &  16 m^2 z_2/(9 t u)
\nonumber
\; ,  
\quad 
\\[2mm]
  n_{C_1}^{(\ep^2)} &\! =\! &
        z_t/t \, n_{C_4}^{(\ep^2)}   \; ,
\nn
\quad 
\\[2mm]
  n_{C_2}^{(\ep^2)} &\! =\! &
        2 z_t/(s\beta^2) n_{C_6}^{(\ep^2)}   \; ,
\nn
\quad  
\\[2mm]
  n_{C_3}^{(\ep^2)} &\! =\! &
        2 t/s \, n_{C_4}^{(\ep^2)}    \; ,
\quad  
\\[2mm]
  n_{C_4}^{(\ep^2)} &\! =\! &
        2 s ( \kappa_{c6} + s^3 - s^2 t )/(D u)  \; ,
\nn
\quad  
\\[2mm]
  n_{C_5}^{(\ep^2)} &\! =\! &
        z_t/(t\beta^2) n_{C_6}^{(\ep^2)}    \; ,
\nn
\quad
\\[2mm]
  n_{C_6}^{(\ep^2)} &\! =\! &
        - 2 s \beta^2 ( \kappa_{c6} - s u z_2 )/(D u)   \; ,
\nn
\quad
\\[2mm]
\nn
   n_{D_1}^{(\ep^2)} &\! =\! &   - t n_{C_6}^{(\ep^2)}  \; ,
\qquad
   n_{D_2}^{(\ep^2)} =   - t n_{C_4}^{(\ep^2)}  \;   .
\ea


\end{document}